\documentclass[twocolumn,trackchanges]{aastex701}
\usepackage{tabularx}
\usepackage{graphicx}
\usepackage[export]{adjustbox}

\graphicspath{{./}{figures/}}

\submitjournal{ApJL}

\shortauthors{Baweja et al.: Co-existence of longitudinal and transverse oscillations in plumes}

\begin{document}

\title{Co-existence of longitudinal and transverse oscillations in polar plumes observed with Solar Orbiter/EUI}

\correspondingauthor{Vaibhav Pant}
\email[show]{vaibhav.pant@aries.res.in, vaibhavpant55@gmail.com}
\author[orcid=0000-0001-7816-1857]{Upasna Baweja}
\affiliation{Aryabhatta Research Institute of Observational Sciences, 263001, Nainital, India}
\email{upasnabaweja.ub@gmail.com}
\affiliation{Department of Applied Physics, Mahatma Jyotiba Phule Rohilkhand University, Bareilly- 243006, Uttar Pradesh, India}
\author[orcid=0000-0002-6954-2276]{Vaibhav Pant}
\affiliation{Aryabhatta Research Institute of Observational Sciences, 263001, Nainital, India}
\email{vaibhav.pant@aries.res.in, vaibhavpant55@gmail.com}
\author[orcid=0000-0002-0735-4501]{S. Krishna Prasad}
\affiliation{Aryabhatta Research Institute of Observational Sciences, 263001, Nainital, India}
\email{krishna.prasad@aries.res.in}
\author[orcid=0000-0001-9035-3245]{Arpit Kumar Shrivastav}
% \affiliation{Aryabhatta Research Institute of Observational Sciences, 263001, Nainital, India}
\affiliation{Southwest Research Institute, Boulder, CO 80302, USA}
\email{arpit.shrivastav@swri.org}
% \author[]{Tom Van Doorsselaere}
\author[orcid=0000-0001-9628-4113]{Tom Van Doorsselaere}
\affiliation{Centre for mathematical Plasma-Astrophysics, Department of Mathematics, KU Leuven, Celestijnenlaan 200B Bus 2400, 3001, Leuven, Belgium}
\email{tom.vandoorsselaere@kuleuven.be}
\author[orcid=0000-0001-9398-2063]{Nancy Narang}
\affiliation{Solar-Terrestrial Centre of Excellence--SIDC, Royal Observatory of Belgium, Ringlaan -3- Av. Circulaire, 1180 Brussels, Belgium}
\email{nancy.narang@oma.be}
\author[orcid=0000-0002-5022-4534]{Cis Verbeeck}
\affiliation{Solar-Terrestrial Centre of Excellence--SIDC, Royal Observatory of Belgium, Ringlaan -3- Av. Circulaire, 1180 Brussels, Belgium}
\email{francis.verbeeck@oma.be}
\author{M. Saleem Khan}
\affiliation{Department of Applied Physics, Mahatma Jyotiba Phule Rohilkhand University, Bareilly- 243006, Uttar Pradesh, India}
\email{}
\author[orcid=0000-0003-4052-9462]{David Berghmans}
\affiliation{Solar-Terrestrial Centre of Excellence--SIDC, Royal Observatory of Belgium, Ringlaan -3- Av. Circulaire, 1180 Brussels, Belgium}
\email{david.berghmans@oma.be}
\begin{abstract}
Magnetohydrodynamic (MHD) waves play a key role in heating the solar corona and driving the solar wind. Recent observations have shown the presence of slow magneto-acoustic and Alfvénic waves in polar plumes and inter-plumes. However, a complete understanding of wave dynamics in the polar regions has long been limited by the lack of simultaneous, high-resolution observations. In this study, we utilize high spatial (210 km per pixel) and high cadence (5s) dataset from the Extreme Ultraviolet Imager (EUI) aboard Solar Orbiter, acquired on 14 September 2021. Our findings reveal the simultaneous presence of slow magneto-acoustic and Alfvénic waves within the same polar plumes. For slow magneto-acoustic waves, the amplitudes of propagating disturbances are 1.4 to 3.2$\%$ of background intensity, with periodicities of 9 min, and the projected speed of these disturbances ranges between 115 to 125 kms$^{-1}$. The corresponding electron temperature in plumes ranges between 0.58 and 0.69 MK. The damping length of these propagating disturbances for five plumes is $\approx$2.4 to 7.1 Mm. The propagating disturbances are also detected in the fine-scale substructures within the plumes. Alfvénic waves, on the other hand, are detected with average displacement amplitude, periodicity, and velocity amplitudes of 165$\pm$82 km, 93$\pm$39 s, and 12$\pm$7 kms$^{-1}$ respectively. The ranges for displacement amplitude, period, and velocity amplitude are 50-600 km, 50-250 s, and 3-32 kms$^{-1}$ respectively. These results mark the first demonstration of Solar Orbiter/EUI's ability to simultaneously detect both slow magneto-acoustic and Alfvénic wave modes extending up to 20 Mm in polar plumes.
\end{abstract}
\keywords{Plumes, Solar corona, MHD waves, Extreme Ultraviolet emission}

\section{Introduction} \label{sec:intro} 
Polar plumes are extended bright narrow structures, observed in the white light and extreme-ultraviolet wavelengths \citep{2015LRSP...12....7P}. These structures subtend an angle of roughly 2$^{\circ}$ relative to the Sun center at low latitude and exhibit super-radial expansion \citep{1997SoPh..175..393D,2001ApJ...560..490D}. These are magnetically open structures. The regions in between these plumes are referred to as inter-plumes. Numerous studies have reported the presence of magnetohydrodynamic (MHD) waves in the off-limb regions of the polar coronal holes (see review by \cite{2021SSRv..217...76B}).

Due to the longer life span of the plumes, they became common features of study during the Solar and Heliospheric Observatory era \citep[SoHO;][]{1995SoPh..162....1D}. \cite{1998ApJ...501L.217D} reported the first detection of the propagating magnetohydrodynamic (MHD) waves in these open coronal structures using SOHO/EIT \citep{1995Delab}. They detected outward propagating features with a speed of 75-150 km/s, and the periodicities of 10-15 minutes. These wave trains are interpreted as propagating slow-mode acoustic waves/compressional waves \citep{2011A&A...528L...4K, 2017ApJ...847L...5P}. Using a two-dimensional MHD model, \cite{2000AdSpR..25.1909O} confirmed these observational findings as slow magneto-acoustic waves. Subsequent modelling efforts by \cite{1999ApJ...514..441O}  and \cite{2000ApJ...533.1071O} revealed that these waves experience nonlinear dissipation. Observations from Ultraviolet Coronagraph Spectrometer \citep[UVCS;][]{1995SoPh..162..313K} onboard SoHO detected these waves farther away from the limb, with periodicities of 6-10 min and sometimes, extending up to 25 minutes \citep{1997ApJ...491L.111O,2000ApJ...529..592O,2000SoPh..196...63B}. In interplume regions, these waves have 20-50 min periodicity up to a height $\leq$20 Mm \citep{2001A&A...377..691B}. These propagating disturbances are generally attributed to the outflows triggered by small-scale reconnection events, such as jets or spicules \citep{2015ApJ...807...71P, 2015ApJ...809L..17J}. \cite{2010ApJ...722.1013D} suggested these quasi-periodic disturbances in the solar corona as the signatures of chromospheric jets that undergo rapid heating to coronal temperatures. Recent observations further indicate that jets within the picoflare energy range, occurring in the coronal hole, may serve as the progenitors of both fast and Alfvénic slow solar winds \citep{2025A&A...694A..71C}. However, contrasting results by \cite{2015ApJ...815L..16S, 2024A&A...689A.135S} suggest that the spicules and associated propagating disturbances may originate from a common physical mechanism.

In addition to slow magneto-acoustic waves, various other MHD wave modes have also been detected in the solar atmosphere \citep{2007SoPh..246....3B, 2008ApJ...676L..73V}. Direct signatures of the kink waves were first reported in the coronal loops \citep{1999ApJ...520..880A,1999Sci...285..862N} using Transition Region And Coronal Explorer \citep[TRACE;][]{1998SoPh..183...29H}.
Alfvénic motions, on the other hand, remained undetected until the high-resolution observations of the chromosphere with Hinode/Solar Optical Telescope \citep{2007Sci...318.1574D} and corona with unique imaging spectroscopy with the Coronal Multichannel Polarimeter \citep[CoMP;][]{2008SoPh..247..411T} \citep{2007Sci...317.1192T,2015NatCo...6.7813M} was made available. Earlier studies investigating the propagating Alfvénic waves in coronal holes relied on nonthermal broadening of spectral lines as an indirect proxy \citep{1991ApJ...375..789S,1998A&A...339..208B,2009A&A...501L..15B,2012ApJ...753...36H,2012ApJ...751..110B}, under the assumption that the excess line broadening was solely due to Alfvénic waves. However, since non-oscillatory mass motions, such as large-scale upflows \citep{2011ApJ...736..130T,2012ApJ...759..144T}, can also contribute to such broadening, direct detection of the transverse waves via imaging offers a more reliable approach over spectroscopic methods. Recently, high-frequency Alfvénic waves have also been detected in the off-limb regions \citep{2025ApJ...982..104M} using high cadence observations from Cryogenic Near-Infrared Spectropolarimeter \citep[Cryo-NIRSP;][]{2023SoPh..298....5F} instrument of the Daniel K. Inouye Solar telescope \citep[DKIST;][]{2020SoPh..295..172R}. Monte Carlo simulations by \cite{2011Natur.475..477M} demonstrated that the transverse wave signatures could be identified in off-limb structures using high-resolution imaging data from Atmospheric Imaging Assembly onboard Solar Dynamic Observatory \citep[SDO;][]{2012SoPh..275....3P} (SDO/AIA). Subsequent investigations by \cite{2014ApJ...790L...2T, 2018ApJ...852...57W,2020ApJ...894...79W} confirmed the presence of transverse waves in plumes using AIA data. Recently, \cite{2025arXiv250113673M} also examined the variation of the wave properties in polar coronal holes over Solar Cycle 24 and found that these wave properties remain relatively consistent over the years. 

% about the limitations in the previous studies
To date, there is just one study where the existence of both these wave modes has been studied in the solar coronal plumes \citep{2015ApJ...806..273L} in addition to the study involving off-limb observations of the active region fan loops \citep{2013A&A...556A.124T}. The authors used imaging observations from AIA/SDO and Doppler velocity observations from CoMP to study the slow compressive waves and Alfvénic waves, respectively. Although the pixel scale within both these datasets is different (4.35 arcsec for CoMP and 0.6 arcsec for AIA), they still managed to observe the transverse and longitudinal oscillations in the polar coronal plumes. 

High-resolution observations from the Extreme Ultraviolet Imager (EUI, \citealt{2020A&A...642A...8R}) onboard Solar Orbiter \citep{2020Muller} have significantly enhanced the detection of transverse and longitudinal oscillations in various coronal structures \citep{2023Petrova, 2024Meadowcroft,2024Shrivastav, 2025Shrivastav}. The primary objective of this letter is to investigate the simultaneous presence of Alfvénic and slow magneto-acoustic waves in the same magnetic structures (coronal plumes) using high-resolution and high-cadence imaging observations from the EUI observations. Following an outline of the data used in this analysis (Section \ref{sec:Observations}), the method to detect the slow-magnetoacoustic waves and transverse waves in the coronal plumes, along with their results, is discussed in Section \ref{sec:Analysis}. A discussion of the results can be found in Section \ref{sec:Summary}. 
\begin{figure*}
    \centering
    \includegraphics[width=0.45\linewidth]{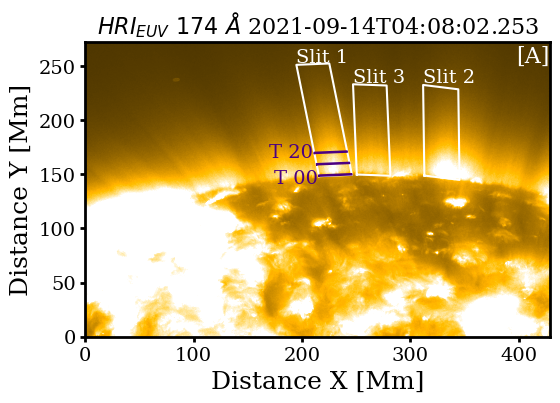}
    \includegraphics[width=0.45\linewidth]{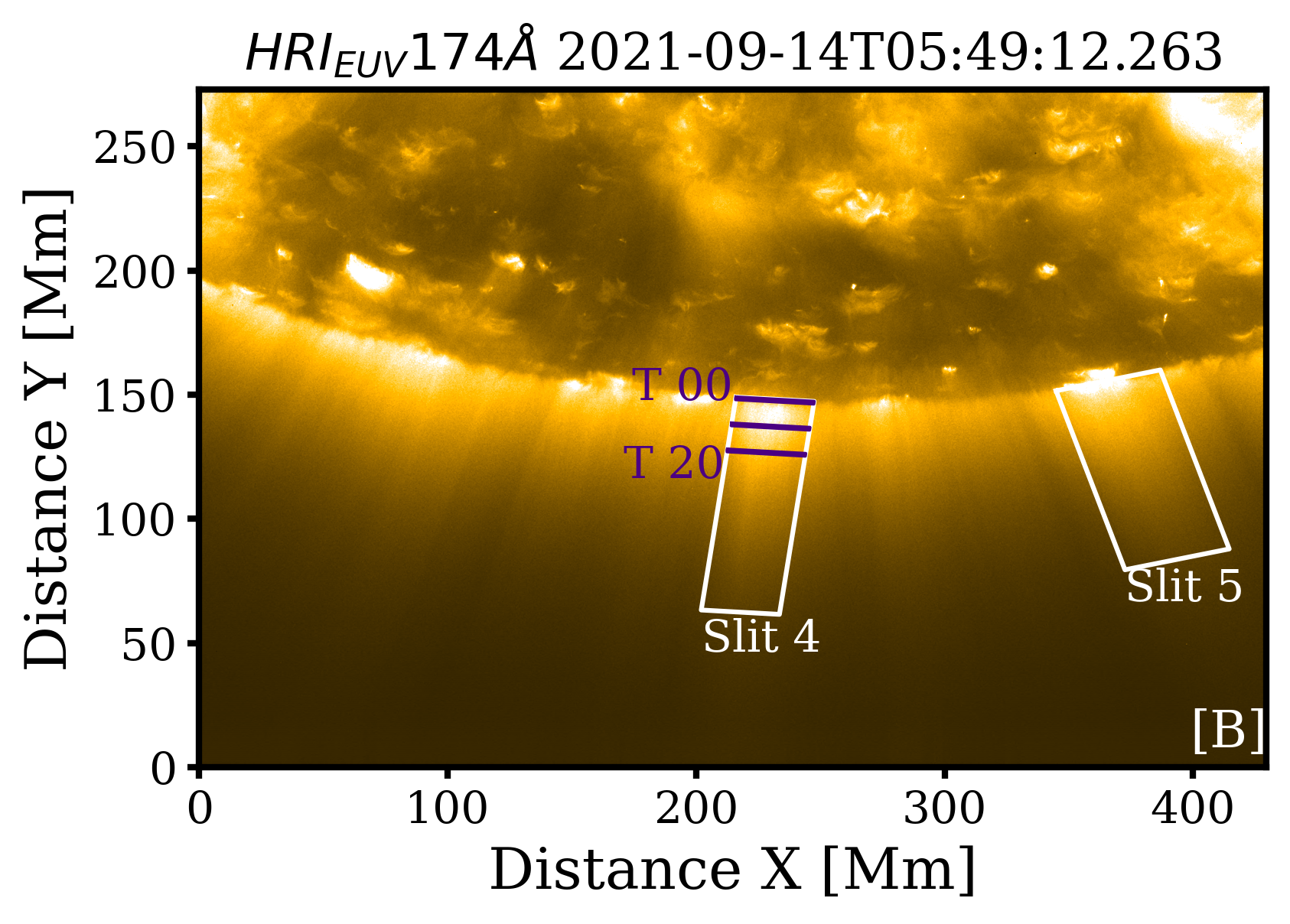}
    \includegraphics[width=0.45\linewidth]{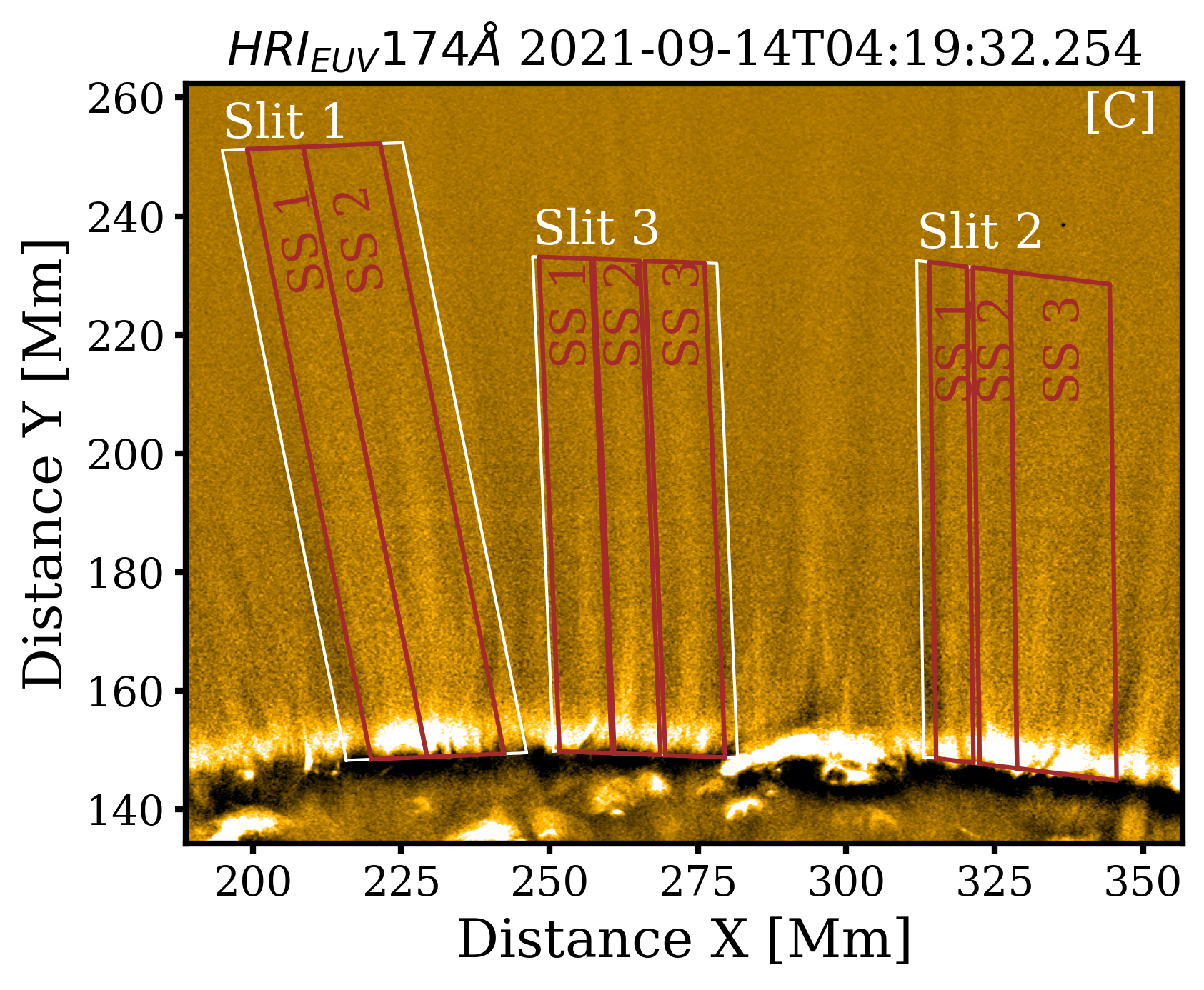}
    \includegraphics[width=0.45\linewidth]{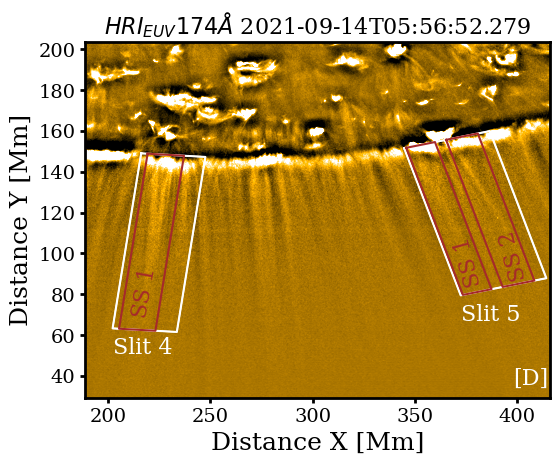}
    \caption{Selected plumes: Panel [A] and [B] represent Dataset 1 and Dataset 2, respectively, as seen by HRI$_{EUV}$ on Solar Orbiter. White rectangles represent the artificial slits on the plumes selected for the analysis, following intensity contours. The indigo-colored rectangles are transverse slits at different heights, where T00 and T20 are the first and 20$^\mathrm{th}$ transverse slits at heights $\approx$3 and 24 Mm above the photospheric limb. Panels [C] and [D] show the zoomed-in, enhanced view of panels [A] and [B], respectively. The brown-colored slits represent the vertical slits on the fine-scale substructures (SS). An animation of panels [C] and [D],illustrating the selected plumes (white contours) along with the selected fine-scale substructures (brown contours) in both Datasets 1 and 2,} is available online.
    \label{fig:datasets}
\end{figure*}
\section{Instrument and Data} \label{sec:Observations}
The observational dataset analyzed in this study is obtained from the High-Resolution EUV Imager (HRIEUV) aboard EUI. The HRIEUV passband centered at 174\,\AA\ captures the emission from the solar plasma at a temperature of approximately 1 MK, with a plate scale of 0.49 arcsec/pixel. The datasets analyzed in this study were recorded on 14 September 2021, which captured polar coronal holes. The first dataset, dataset-1, corresponds to the north polar coronal hole, the imaging sequence of which is taken from 04:08:02 - 04:26:43 UT, spanning approximately 19 minutes. The second dataset, dataset-2, covers the south polar coronal hole from 05:53:02 - 06:11:43 UT. Panels [A] and [B] of Figure \ref{fig:datasets} represent the context image of these datasets. During the time of observation, the Solar Orbiter was positioned at 0.59 au from the Sun, resulting in an effective pixel resolution of 210 km. The imaging sequence for both datasets was taken with an exposure time of 2.8 s and a cadence of 5 s. Both these datasets were also employed in the investigation of transverse decayless waves in short loops within coronal holes and the quiet Sun by \cite{2024Shrivastav}. For the present study, we utilized the level-2 FITS files from EUI data release-5 \citep{euidatarelease5} available at the EUI website\footnote{\url{https://www.sidc.be/EUI/data/}} and SOAR (Solar Orbiter Archive\footnote{\url{https://soar.esac.esa.int/soar/}}). These datasets were acquired as a part of Solar Orbiter Observing Plan (SOOP) ``\textnormal{L\_SMALL\_HRES\_HCAD\_Slow-Wind-Connection}'' \citep{2020A&A...642A...3Z}.  

% The selected regions are located at the solar north and south poles within large coronal holes, where plumes originate. 
To eliminate the effect of the spacecraft jitter from these datasets, a cross-correlation-based image-alignment technique is employed using fg$\_$rigidalign.pro procedure in IDL (Interactive Data Language). This alignment is repeated 5 to 6 times to accurately co-align all the images with the first image of the sequence. Further, to distinguish the plume and interplume locations, the intensity contours are plotted \citep{2011A&A...528L...4K}. Five distinct plumes, three from dataset-1 and two from dataset-2 are selected for investigation and are highlighted as white rectangles in Figure \ref{fig:datasets}. The width of the selected plumes is listed in Table \ref{tab:propagationspeed}. To enhance the visibility of the fine-scale substructures within these plumes, the datasets are subject to spatial filtering procedure, in which each image in the sequence has an 11 $\times$ 11 Mm$^2$ boxcar-smoothed image subtracted from itself. This will preserve finer features while removing those at scales larger than 11 Mm. The resulting sequence is further smoothed over five time frames to suppress the frame-to-frame variation in intensity \citep{2014ApJ...790L...2T}. A zoomed version of these fine-scale substructures within the selected plumes is illustrated in Panels [C] and [D] of Figure \ref{fig:datasets}. This processed image sequence, obtained after removing all the larger structures, is further referred to as processed datacube and is used only for the analysis of Alfvénic waves. In contrast, for the investigation of slow magneto-acoustic waves, the sequence obtained after removing the jitter, but without precessing, is used and is referred to as the original datacube.

\section{Data analysis and Results} \label{sec:Analysis}
% Further details of the dataset are provided in Table \ref{tab:Data}.
\subsection{Investigation of slow magneto-acoustic waves} \label{sec:slowmodedetection}
To investigate slow magneto-acoustic waves in coronal plumes, time-distance (TD) maps of the selected vertical slits (as shown in Figure \ref{fig:datasets}) are extracted from the original datacube. The propagating disturbances appear as alternating bright and dark ridges with finite slopes in the intensity TD maps, indicating compression and rarefaction of the wave propagating within the plume structure. The time-averaged intensities of these extracted TD maps are used as the background is subtracted from the TD maps to enhance the ridge visibility. Panels [A] to [E] of Figure \ref{fig:vertical_slits} represent these background-subtracted TD maps for all five selected plumes. Note that the bottom of these maps corresponds to the height 3.15 Mm above the solar limb. However, only a few ridges are visible due to the short dataset duration (19 minutes) as the periodicity of these slow magneto-acoustic waves in plumes is tens of minutes \citep{2021SSRv..217...76B}.

\begin{table*}
    \centering
    \begin{tabular*}{\textwidth}{@{\extracolsep{\fill}}|l|cccc|cccc|cccc|}
    \hline
    Plumes & \multicolumn{4}{c|}{Width (Mm)} & \multicolumn{4}{c|}{Propagation Speed (km/s)} & \multicolumn{4}{c|}{}  \\
    \cline{2-9}
    & WP & SS1 & SS2 & SS3 & WP & SS1 & SS2 & SS3 & $A_{0l}$ (\%) & $L_d$ (Mm) & $P_l$(min)& $T$(MK) \\
    \hline
    Slit 1 & 30.4 & 13.0 & 9.4 & -- &  115$\pm$2 &  131$\pm$2 &  105$\pm$2 & -- & 3.2 &  5.00$\pm$0.10 & 9.4& 0.58$\pm$0.02 \\
    Slit 2 & 32.5 & 6.3 & 6.3 & 16.8 &  116$\pm$6 &  107$\pm$4 &  93$\pm$2 &  72$\pm$3 & 1.9 &  2.45$\pm$0.20 & --& 0.59$\pm$0.06\\
    Slit 3 & 31.1 & 8.8 & 7.8 & 10.1 &  125$\pm$4 &  128$\pm$3 &  123$\pm$3 &  137$\pm$3 & 1.7 &  7.16$\pm$0.13 & 9.4& 0.69$\pm$0.04\\
    Slit 4 & 31.5 & 17.8 & -- & -- &  125$\pm$3 &  112$\pm$2 & -- & -- & 1.4 &  6.02$\pm$0.20 & --& 0.69$\pm$0.03\\
    Slit 5 & 42.0 & 14.1 & 15.5 & -- &  116$\pm$2 &  96$\pm$4 &  86$\pm$2 & -- & 2.7 &  3.52$\pm$0.08 & --& 0.59$\pm$0.02\\
    \hline
    \end{tabular*}
    \caption{Propagation speeds of slow magneto-acoustic waves in the selected plumes and their fine-scale substructures (SS), along with their widths, relative amplitudes ($A_{0l}$), damping lengths ($L_d$), longitudinal periods ($P_l$),  and electron temperature ($T$)}. (WP corresponds to the whole plume)
    \label{tab:propagationspeed}
\end{table*}

\subsubsection{Determining wave parameters of slow magneto-acoustic waves}
(i) Speed of propagating disturbances
 
\begin{figure*}
    \centering
    \includegraphics[width=0.17\linewidth]{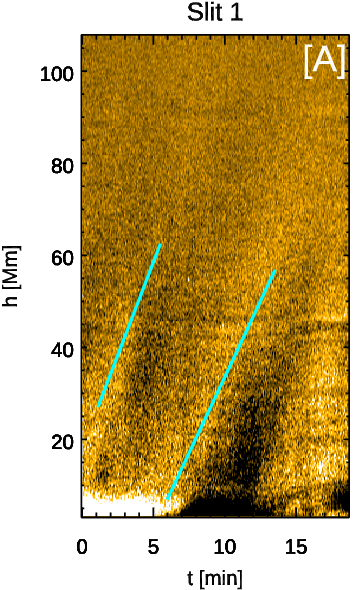}
    \includegraphics[width=0.17\linewidth]{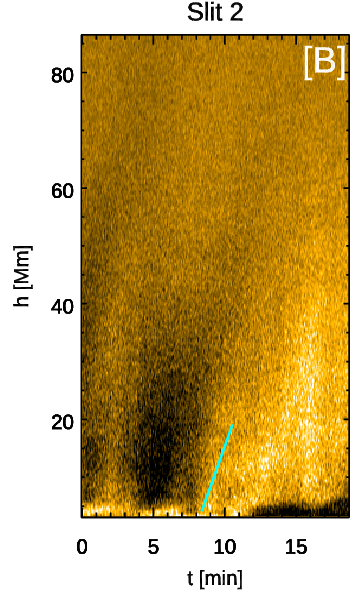}
    \includegraphics[width=0.17\linewidth]{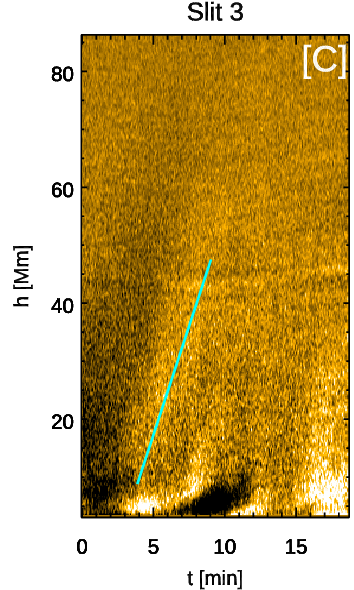}
    \includegraphics[width=0.17\linewidth]{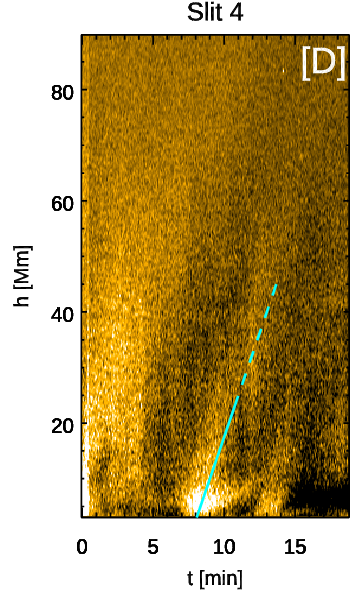}
    \includegraphics[width=0.17\linewidth]{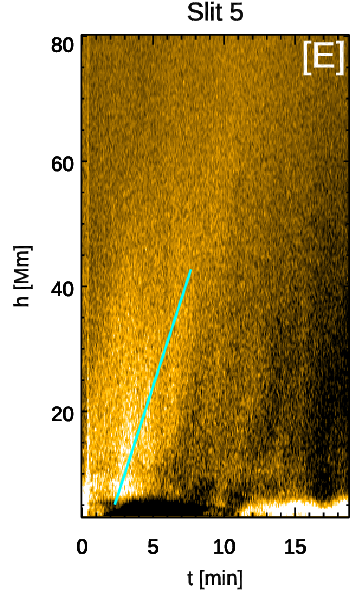}
    \includegraphics[width=0.17\linewidth]{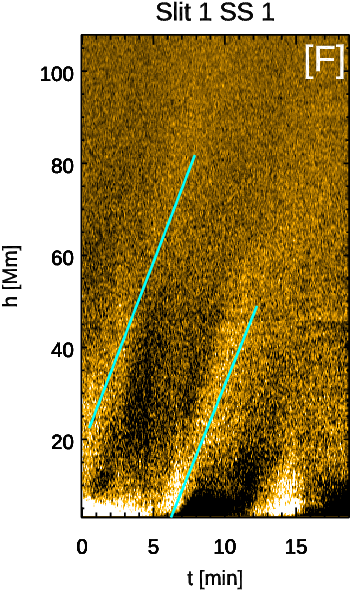}
    \includegraphics[width=0.17\linewidth]{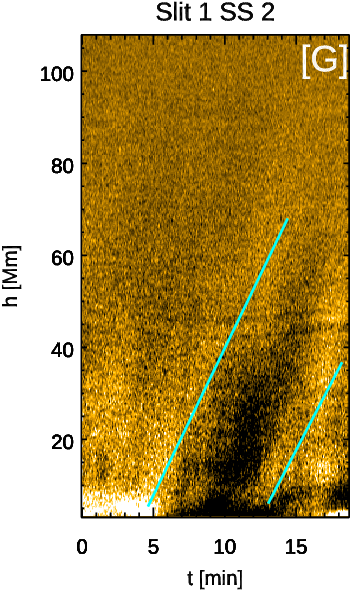}
    \includegraphics[width=0.17\linewidth]{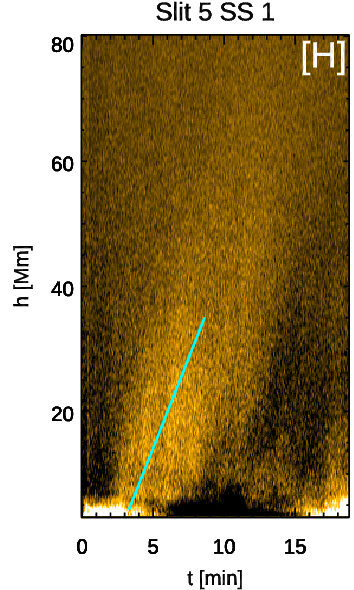}
    \includegraphics[width=0.17\linewidth]{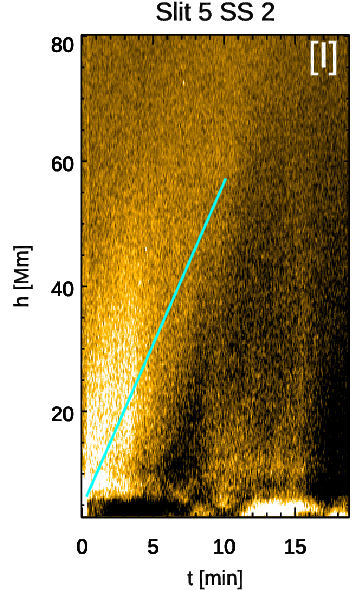}
    \caption{Propagating Disturbances: Panels [A]- [E] show the extracted time-distance (TD) maps for the selected plumes. Slit numbers are mentioned at the top of each TD map. The cyan lines are fitted on the ridges depicting the speed of the propagating disturbances in the slits. Panels [F] and [G] represent the TD maps and the fitted ridges of the two fine-scale substructures (SS 1 and SS 2) corresponding to Slit 1. Panels [H] and [I] represent the same for fine-scale substructures of Slit 5. }
    \label{fig:vertical_slits}
\end{figure*}

The ridge slopes provide projected propagation speeds of these compressive waves. To quantify this, a clean ridge is extracted from the background-subtracted TD map, ensuring a sufficiently wide selection to isolate a single ridge. A Gaussian profile is fitted at each spatial distance along the ridge to identify the local maxima, and the resulting peak locations are then used in a linear fit to determine the propagation speed. The uncertainty from the Gaussian fit is incorporated into the linear regression, and the 1-sigma uncertainty from the fit is considered as the uncertainty in the propagation speed. The selected ridges for which the propagation speeds are obtained are highlighted as cyan lines in Figure \ref{fig:vertical_slits}.
Further details on this method are available in \cite{2012SoPh..281...67K}. The propagation speeds of these plumes, along with their uncertainties, are listed in Table \ref{tab:propagationspeed}.
It should be noted that ridges are not clearly identified in the TD maps of slit 2, which is located near the edge of the field of view. Additionally, for slit 4, there appears to be a discontinuity in the ridge after 11 minutes, which is why a dashed line is fitted beyond this. Although Figure \ref{fig:vertical_slits} represents the analysis for one particular width of the vertical slit, the presence of these ridges is confirmed using multiple slit widths. This ensures that a coherent physical mechanism is involved. 
(ii) Amplitude and Damping length

\begin{figure*}
    \centering
    \includegraphics[width=0.3\linewidth]{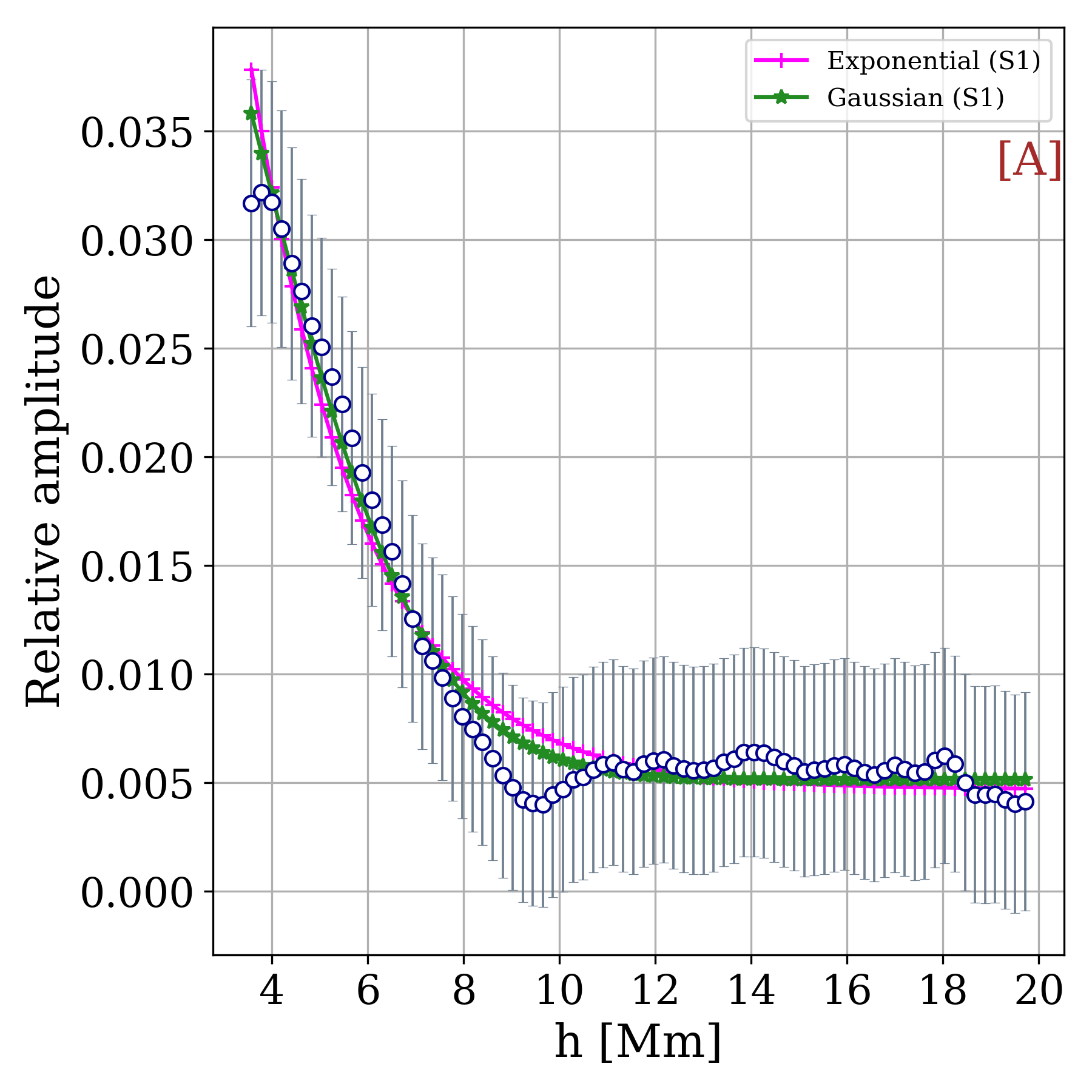}
    \includegraphics[width=0.3\linewidth]{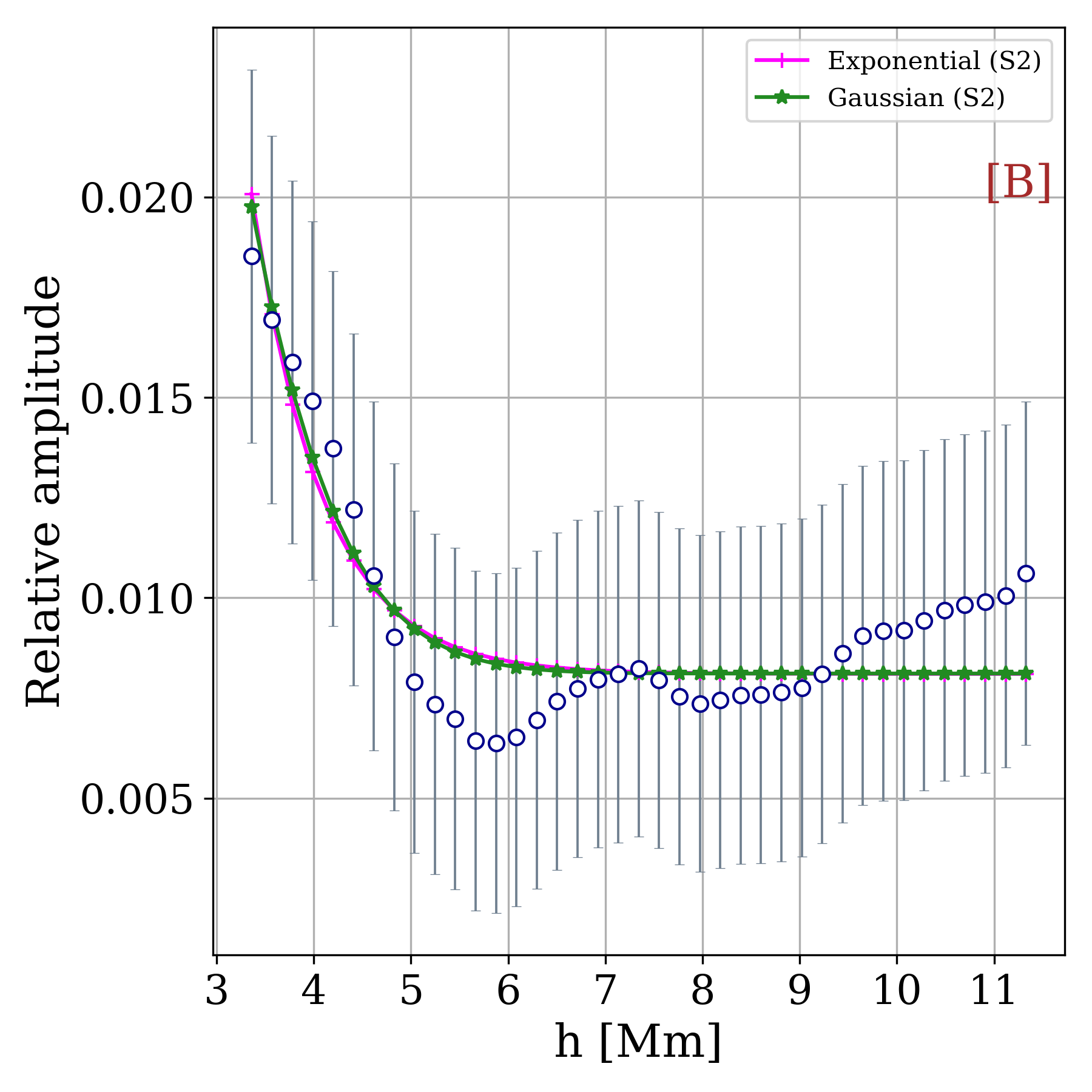}
    \includegraphics[width=0.3\linewidth]{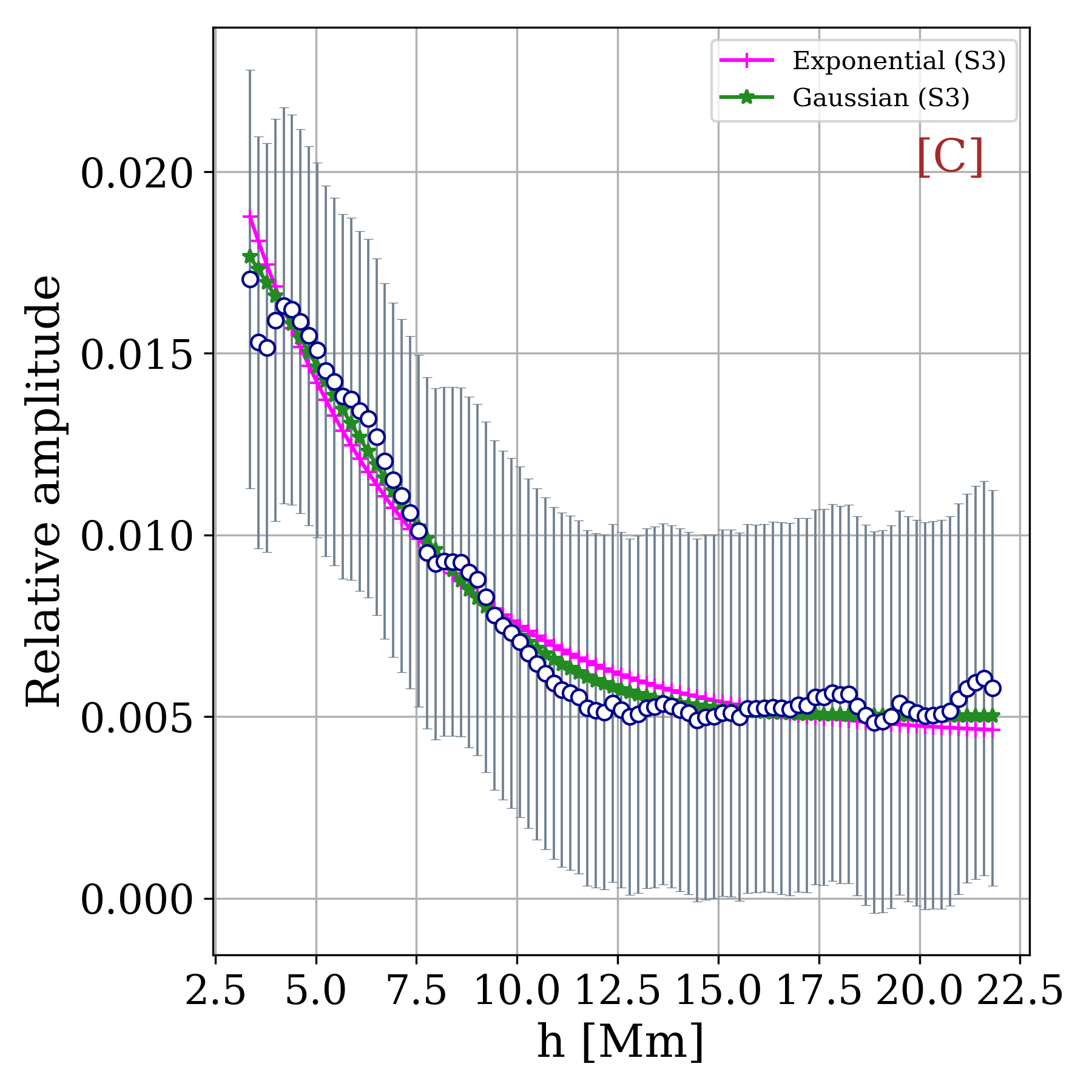}
    \includegraphics[width=0.3\linewidth]{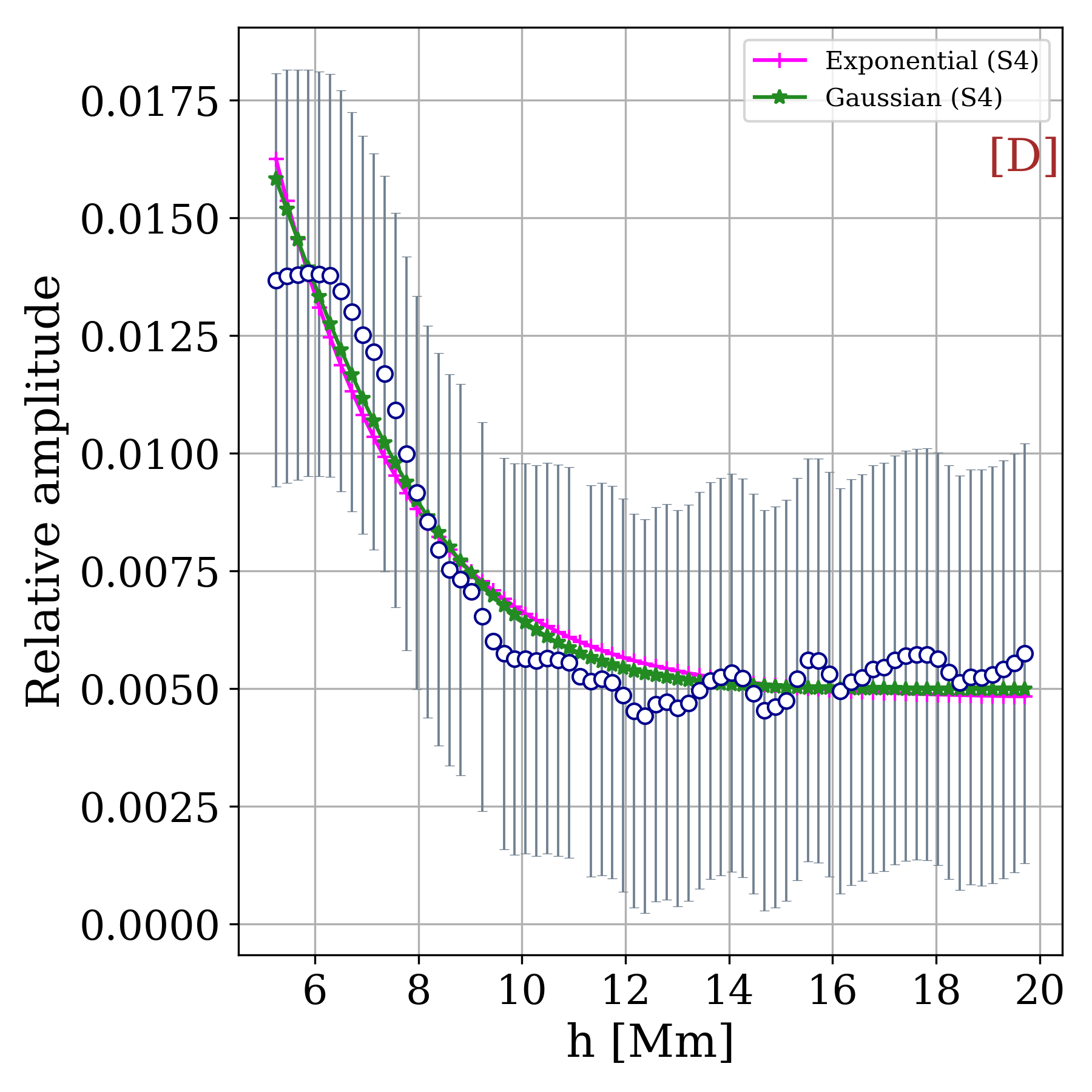}
    \includegraphics[width=0.3\linewidth]{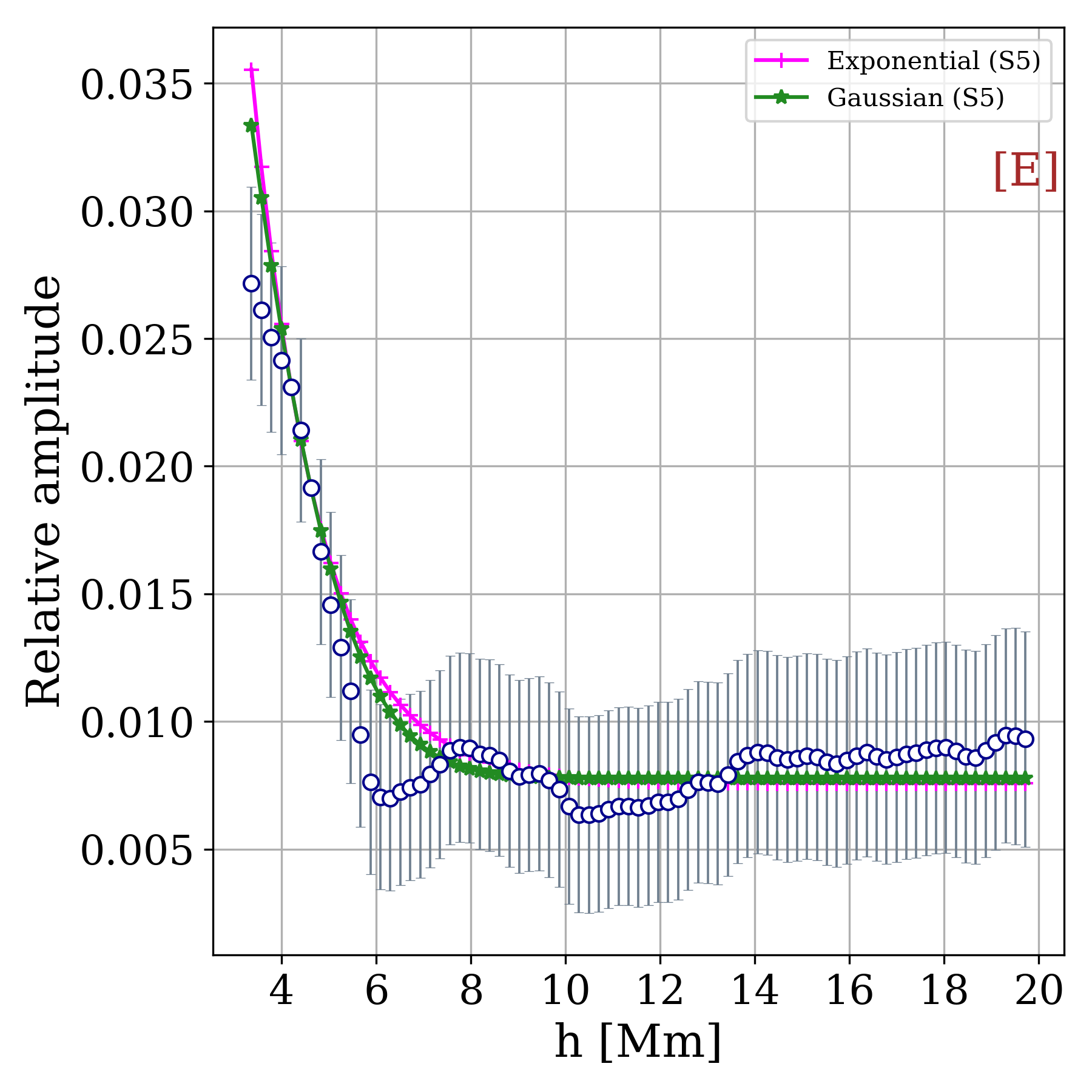}
    
    \includegraphics[width=0.3\linewidth]{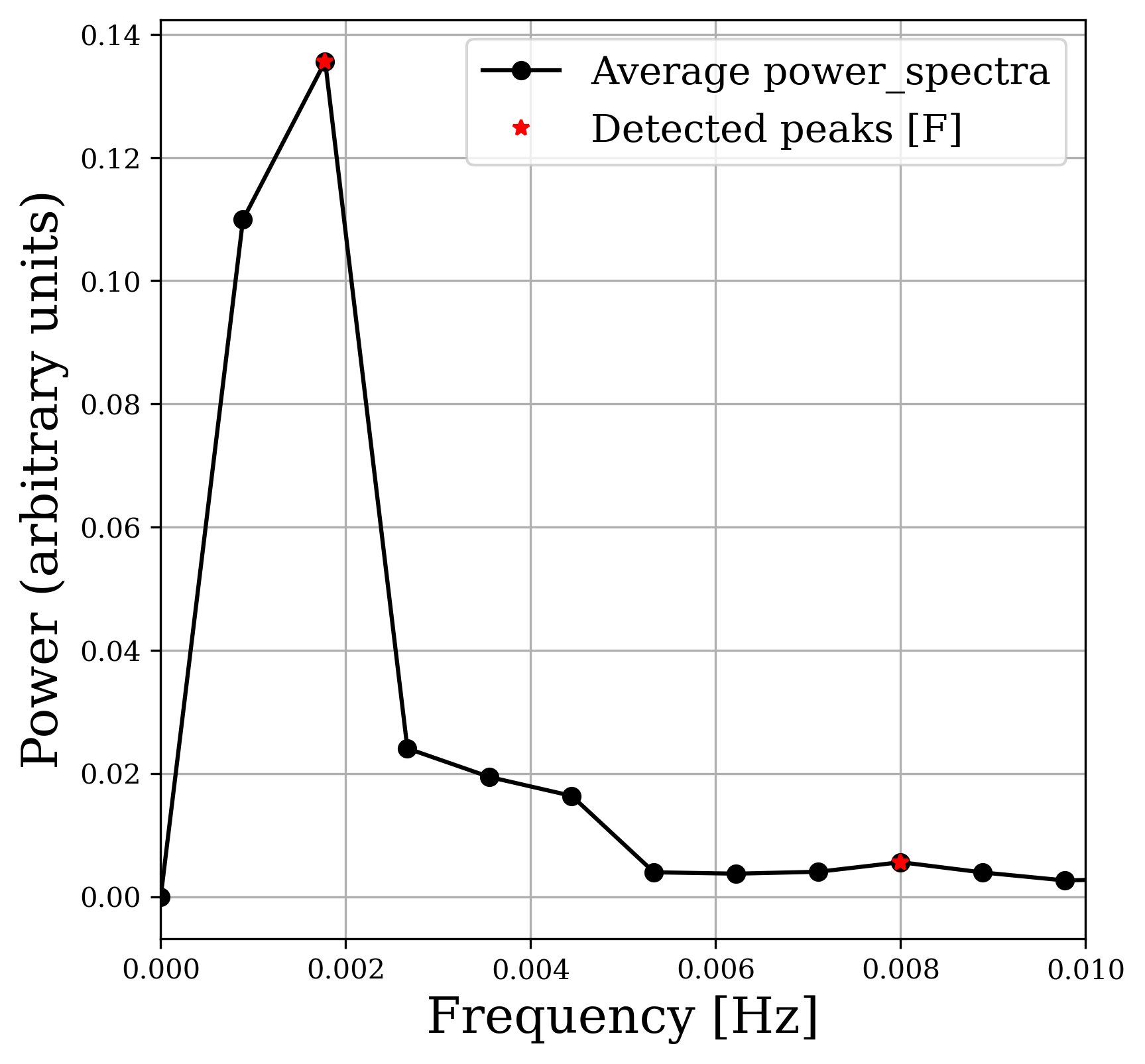}
    \includegraphics[width=0.3\linewidth]{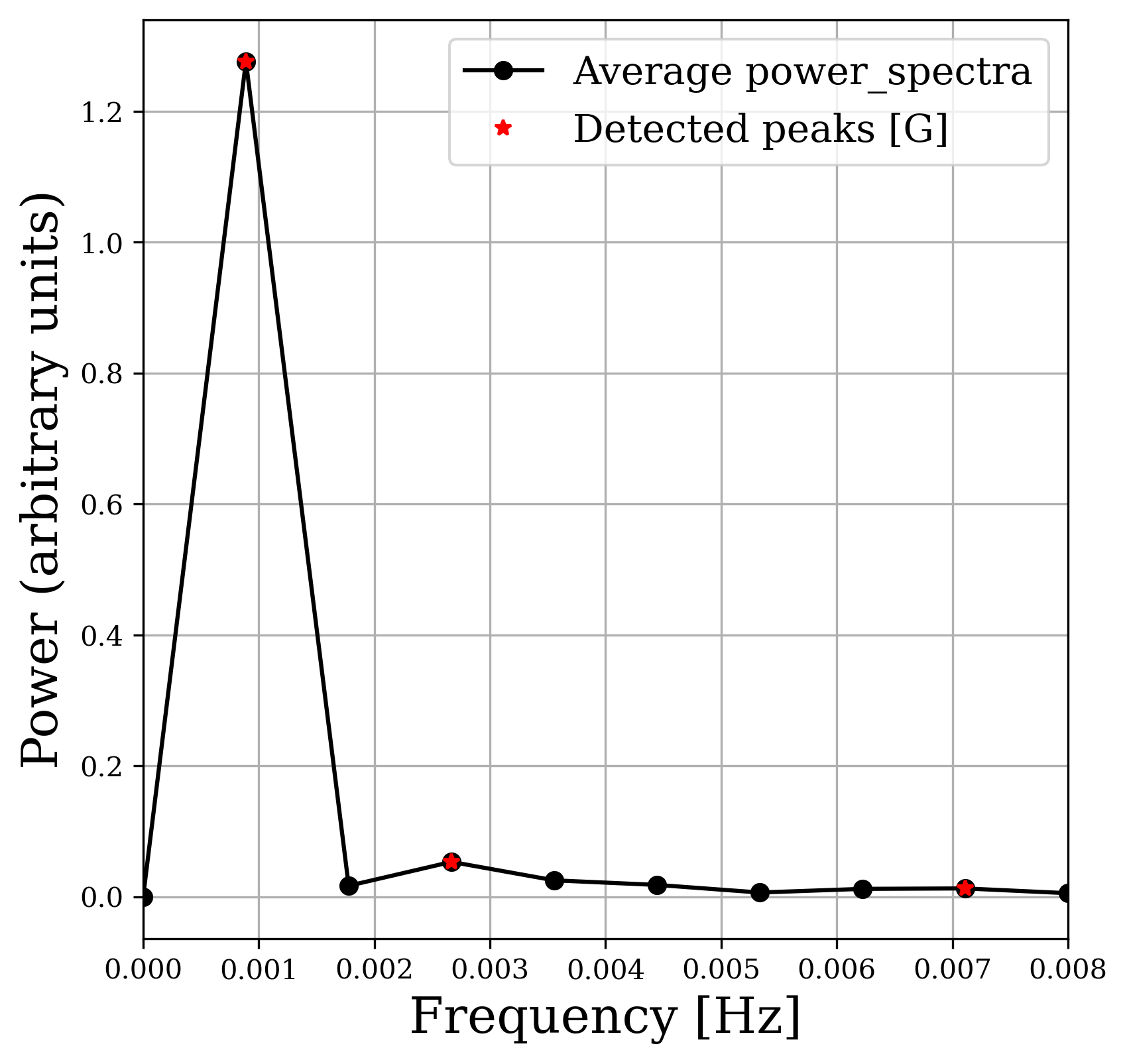}
    \includegraphics[width=0.3\linewidth]{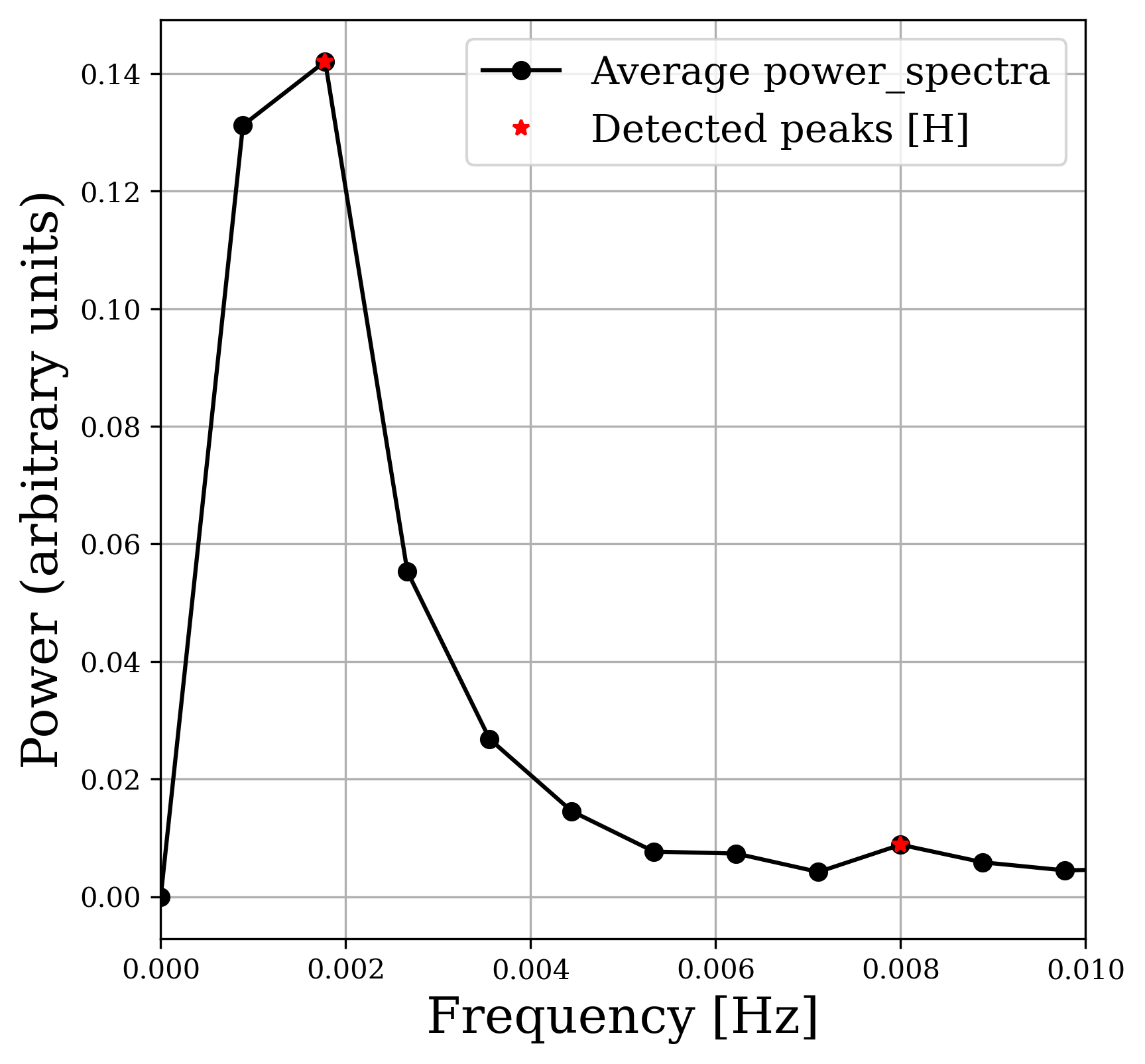}
    \includegraphics[width=0.3\linewidth]{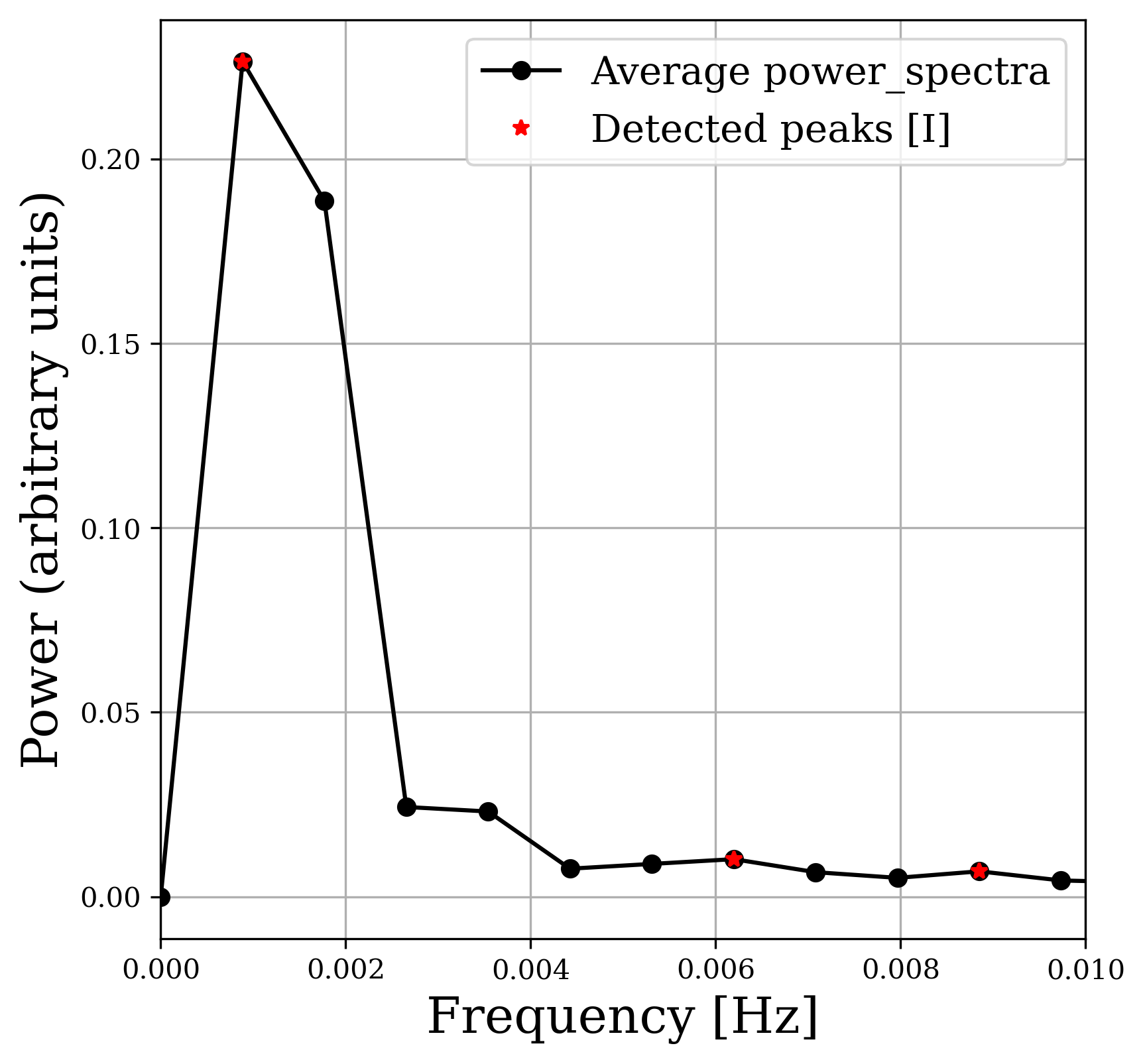}
    \includegraphics[width=0.3\linewidth]{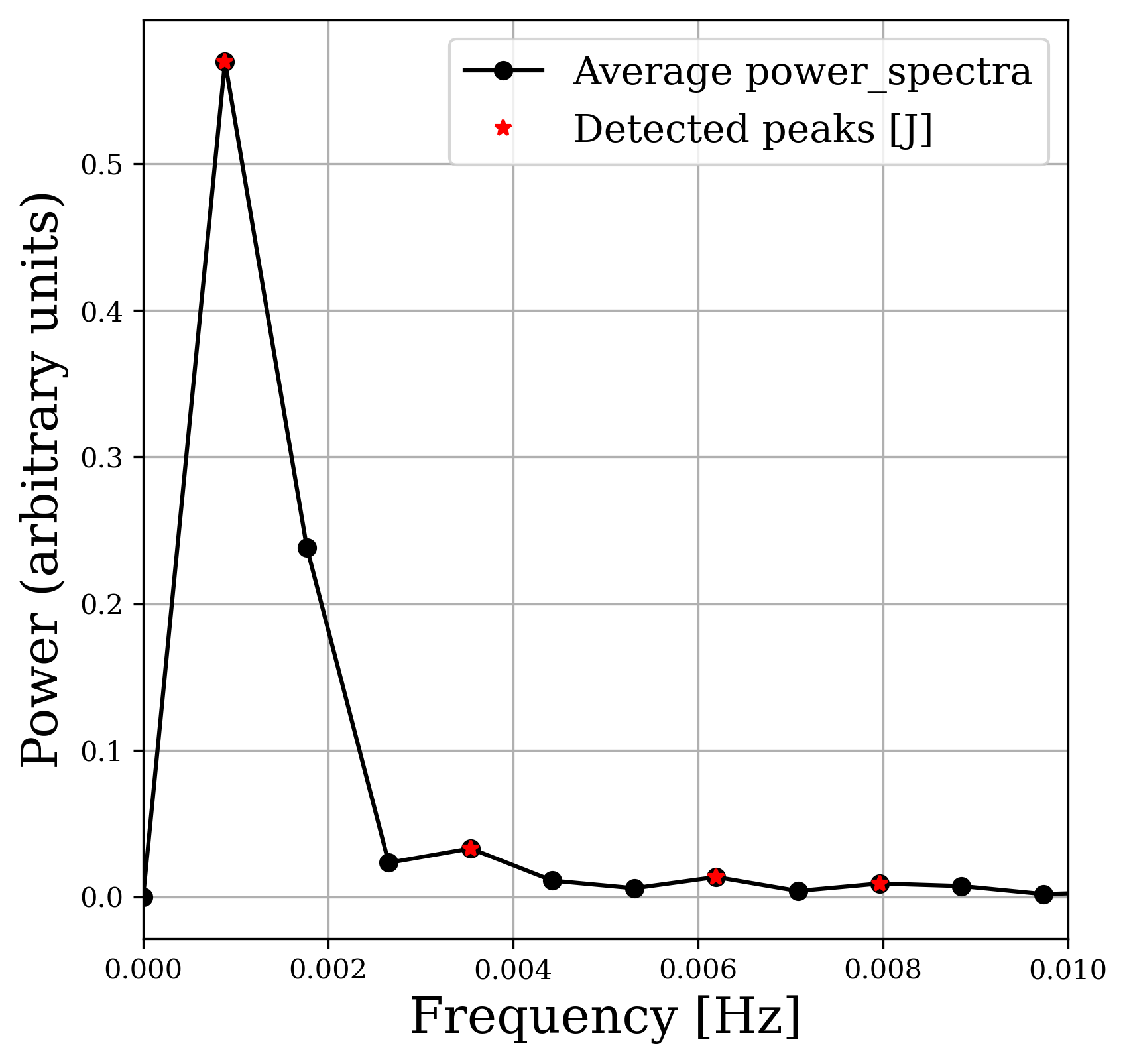}
    \caption{Damping lengths and periodicity: Panels [A]- [E] show the relative amplitudes of the selected plumes as a function of the distance along the slits. Slit numbers are mentioned at the top of each profile. The grey error bars represent the associated uncertainties. The green stars and magenta plus sign represent the Gaussian and exponential fits to the decaying spatial profiles. Panels [F]-[J] represent the average Fourier power spectra at different heights of the different plumes. The red stars represent the dominant frequencies for these plumes. }
    \label{fig:damping_profiles}
\end{figure*}

The background removed TD maps, shown in Panels [A]-[E] of Figure \ref{fig:vertical_slits}, reveal the gradual decrease in the oscillation amplitude with height along the plumes. This spatial decay in amplitude indicates the physical wave dissipation. To study the damping properties of the slow magneto-acoustic waves within the selected plumes, we estimated the damping length using the amplitude tracking method described in \cite{2019FrASS...6...57S, tripathy2025propertiesslowmagnetoacousticwaves}. These background-subtracted TD maps are further normalized. At each spatial location, the oscillation curves are boxcar smoothed to reduce the high-frequency noise in these curves. These smoothed curves can now be considered as sinusoidal curves and the amplitudes ($A_l$) of which at each spatial distance are estimated using \(A_l = \sqrt{2} \sigma \), where $\sigma$ is the standard deviation of the smoothed oscillation curve. The spatial profile reveals the rapid decrease in the relative amplitude of the oscillations with height as in panels [A]-[E] of Figure \ref{fig:damping_profiles}. 

The uncertainty in the imaging intensity at each location is estimated by considering both the readout noise and photon noise (following the methodology of \cite{2023Petrova, 2024Shrivastav}). This uncertainty is further propagated to the background-subtracted and normalized TD maps. The grey colored error bars in each panel [A] -[E] of Figure \ref{fig:damping_profiles} represent the median uncertainties derived from the time series of the normalized TD maps. The spatial profiles of the amplitude are then fitted with an exponentially decaying function \(A_{le}(x) = A_{0l} \exp(\frac{-x}{L_{de}})+C\) and Gaussian decaying function \((A_{lg}(x) = A_{0l} \exp(\frac{-x}{L_{dg}})+C\) to obtain the damping lengths. Here $x$ is the distance along the loop, $L_{de}$ and $L_{dg}$ are the damping lengths obtained using the exponential decaying function and Gaussian decaying function, respectively, $A_{0l}$ and $C$ are the appropriate constants. To isolate the decaying oscillations, the first few initial points are generally neglected, where the amplitudes are increasing. The uncertainty in the damping lengths is the 1-sigma error obtained from the fitting. The amplitude decays are better fitted with the Gaussian decaying function rather than the exponential decaying function, with their reduced chi-square values being less than the chi-square values of the exponential decaying function. This similar behavior was also reported in \cite{2014ApJ...789..118K, 2018ApJ...853..134M}. However, the Gaussian decay fit was better fitted for the amplitude decay with periodicities $\ge$20 minutes, and for the plumes considered in this study, the periodicities of three plumes appear to be greater than the observation time (19 min). Thus, the damping lengths reported in Table \ref{tab:propagationspeed} are the ones obtained from the Gaussian decaying function. 

Notably, the damping lengths $L_{dg}$ obtained are lesser than those obtained earlier in plumes using AIA 171 $\AA$ dataset, which was $\approx$ 23 Mm for periodicity of 9 minutes \citep{2014ApJ...789..118K, 2018ApJ...853..134M}. The reason for the shorter damping lengths obtained in this study is not known, but the spatial profiles after 20 Mm are mostly dominated by noise rather than the actual signal, although the signatures of these propagating disturbances above 20 Mm are visible in the TD maps (Figure \ref{fig:vertical_slits}).  The damping lengths in a few studies \citep{2024Meadowcroft} are reported to be dependent on the observing instrument; on the other hand, some studies contradict this dependence \citep{tripathy2025propertiesslowmagnetoacousticwaves}. The maximum relative amplitudes for each plume are reported as the amplitude $A_{0l}$ of the slow magneto-acoustic waves as compared to the background. These estimated wave parameters ($A_{0l}$ and $L_d$) for each selected plume are reported in Table \ref{tab:propagationspeed}.

(iii) Periodicities of slow magneto-acoustic waves

To estimate the wave period of the slow magneto-acoustic waves, the background-subtracted TD maps are employed. These periods are determined by selecting the peaks in the Fourier power spectra derived from the background-subtracted TD maps, as shown in Panels [F]-[J] of Figure \ref{fig:damping_profiles}. The dominant peaks of the selected plumes are marked with a red star. For each plume, temporal signals at different spatial heights are averaged to obtain the Fourier power spectra. Periods are identified for Slit 1 and Slit 3, both showing dominant periodicity values of 9.4 min. However, for other slits, no distinct periodicities are obtained. As already mentioned in Section \ref{sec:Observations}, the duration of the data itself is just 19 minutes, thus making it difficult to capture low-frequency components for Slits 2, 4, and 5. The peaks in panels [G], [H], and [J] of Figure \ref{fig:damping_profiles} correspond to 18.8 minutes, which is the duration of the dataset under consideration. The periodicities are reported in Table \ref{tab:propagationspeed}. It should be noted that the periodicities reported here are based on the dominant frequencies in the Fourier power spectra and may be less reliable due to the short duration (18.8 minutes) of the dataset. Apart from a 9.4-minute periodicity in two plumes, shorter periodicities of 1.8, 2.6, and 6.25 minutes are also found. Such short-range periodicities in plumes are also observed in previous studies \citep{2014ApJ...789..118K,2018ApJ...853..134M}. These short-range periodicities are perhaps the result of leakage of global p-modes or chromospheric acoustic resonances or acoustic cutoff effect involving impulsive disturbances \citep{2005ApJ...624L..61D, 2011ApJ...728...84B, 2013SoPh..286..405C, 2015ApJ...807...71P, 2018ApJ...853..134M}.

(iv) Electron temperature \\
The electron temperature $T$ in the solar coronal plumes can be inferred by considering the projected propagation (phase) speed of the slow magneto-acoustic waves to coincide with the local sound speed $c_s$. The local sound speed is related to the temperature by the relation \(c_s \approx 151 \sqrt{T}\) \citep{2019mwsa.book.....R}, where $c_s$ is in km s$^{-1}$ and $T$ is in million kelvin (MK). Based on the propagation speed of the slow magneto-acoustic waves in whole plumes, which range from 115 to 125 km s$^{-1}$, the corresponding electron temperature ($T$) values are reported in Table \ref{tab:propagationspeed}. The electron temperature of the selected plumes varies in the range 0.58-0.69 MK. Although the estimated temperature is consistent with the values obtained in earlier studies \citep{1998SSRv...85..371W,1998A&A...336L..90D,1999JGR...104.9753D,2003A&A...398..743D,2006A&A...455..697W,2008ApJ...685.1270L}, they should be considered as the lower limits, since the estimation is based on the projected propagation speed. The errors in the estimated temperatures are obtained by propagating the errors from the projected phase speed and are thus very small.
\subsubsection{Propagating disturbances in fine-scale substructures}
Plumes also contain fine-scale substructures, which are highlighted with brown colored slits in Figure \ref{fig:datasets} panels [C] and [D]. The width of these fine-scale substructures is selected so that they remain within these artificial slits for the whole duration of observations (as can be clearly seen in the animation available). This method differs from the whole plume width selection, which is guided by the intensity contours. Each substructure is individually analyzed, and its corresponding TD maps are extracted (panels [F] to [I] Figure \ref{fig:vertical_slits}), which also exhibit alternating bright and dark ridges. The propagation speed and the width of the selected fine-scale substructures are reported in Table \ref{tab:propagationspeed}. 

For Slit 1, three ridges are distinctly visible in the TD map, as shown in Figure \ref{fig:vertical_slits} panel [A]. Slit 1 has two different fine-scale structures as shown in Panel [C] of Figure \ref{fig:datasets}. The extracted TD maps corresponding to both these fine-scale substructures (SS 1 and SS 2) also exhibit alternate bright and dark ridges as shown in panels [F] and [G] of Figure \ref{fig:vertical_slits}. There are three ridges in the TD maps of each substructure corresponding to Slit 1, out of which two are fitted with cyan lines, as the third one (rightmost ridge starting at time 11 minutes in substructure 1 and leftmost ridge in substructure 2 indicates their presence faintly in the TD maps.) It is evident that the features visible in the full TD map of Slit 1 arise from the superposition of ridges from individual substructures (SS 1 and SS 2). It should be noted that all three ridges originate at the same time in the TD maps of the whole slit and the respective substructures. This implies that a coherent mechanism is involved. 

For the slit 5 (Figure \ref{fig:vertical_slits} panel [E]), a broader ridge is visible in the whole plume TD map. However, when the analysis is performed for the individual fine-scale substructure (as shown in Figure \ref{fig:datasets} panel [D]), distinct ridges indicate their presence in their TD maps (Figure \ref{fig:vertical_slits} panels [H] and [I]).
Additionally, from the TD maps of the whole slit and the fine-scale substructures of Slit 5, the ridge is visible at the beginning of the TD map. 

Furthermore, one can also notice the presence of some thin bright ridges that run parallel to the distance axis in the TD maps of the whole plume or even the fine-scale substructure (Panels [E] and [I] of Figure \ref{fig:vertical_slits}) during the time interval between 0 to 5 minutes and 14 to 16 minutes. This is due to the interference of the fine-scale substructure, which was not considered in substructures 1 and 2, and lies in between them, and appears at the edge of the slit in substructure 2. Thus, such vertical bright ridges are not fitted here. Further details on the fine-scale substructures of slit 2, 3, and 4 can be found in Appendix \ref{AppendixA}.

\subsection{Investigation of Alfvénic waves} \label{sec:fastmodedetection}
To investigate Alfvénic wave signatures, the signal-to-noise ratio of the data is improved by processing the original datacube as mentioned in Section \ref{sec:Observations}. As a result, a processed datacube is obtained in which fine-scale substructures within the same plume become more distinguishable, as shown in panels [C] and [D] of Figure \ref{fig:datasets}. Within the selected coronal plumes, transverse slits are placed inside the slits. These transverse slits have a fixed width of 5 pixels (1.05 Mm) and are contiguous, meaning there is no separation between them. TD maps are generated for these transverse slits in each plume, and transverse oscillations are manually identified.
\subsubsection{Determining wave parameters of Alfvénic waves}
\begin{figure*}
    \centering
    \includegraphics[height = 9cm]{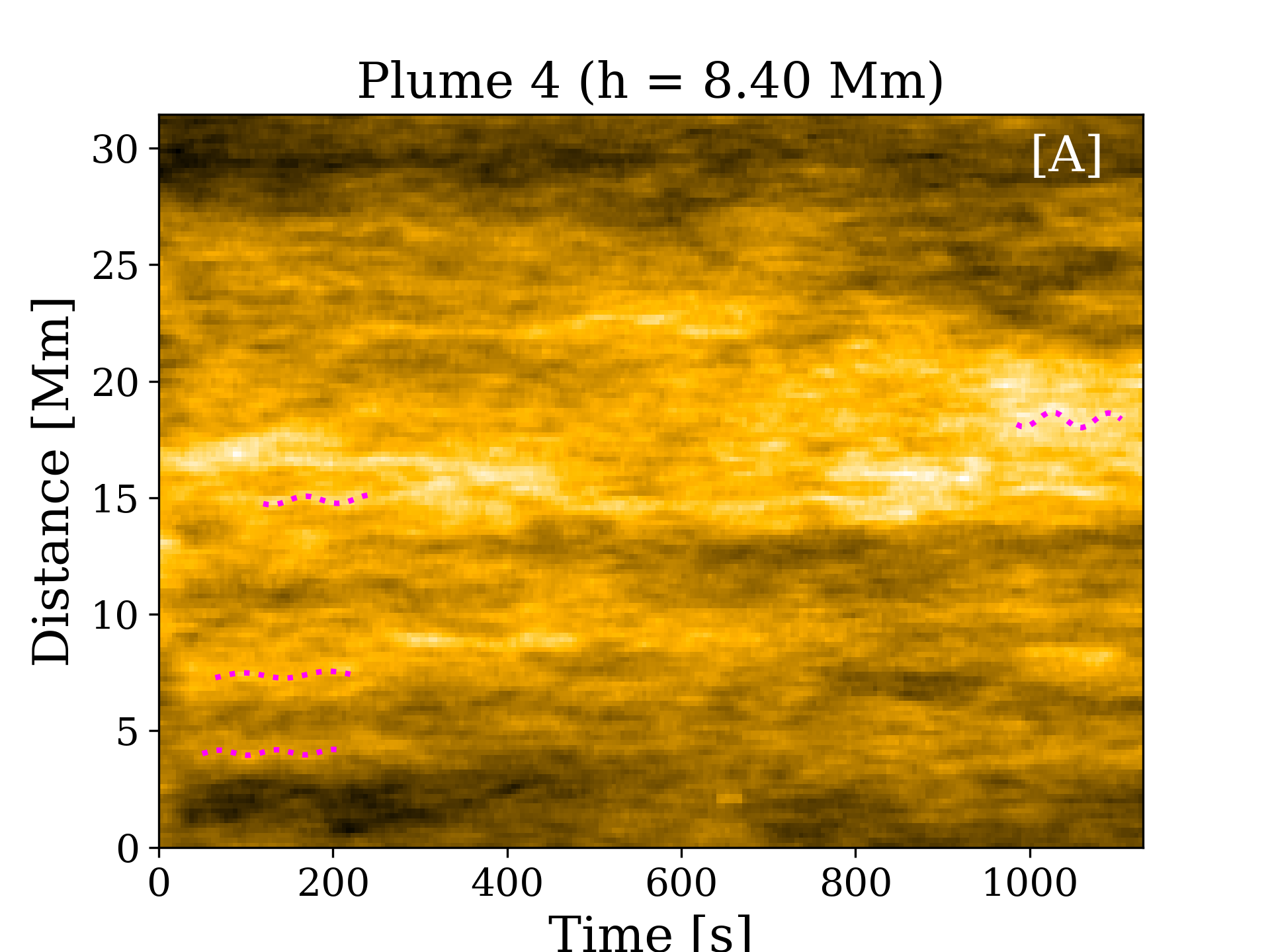}
    % \hspace{0.4cm}
    \includegraphics[width=0.40\textwidth]{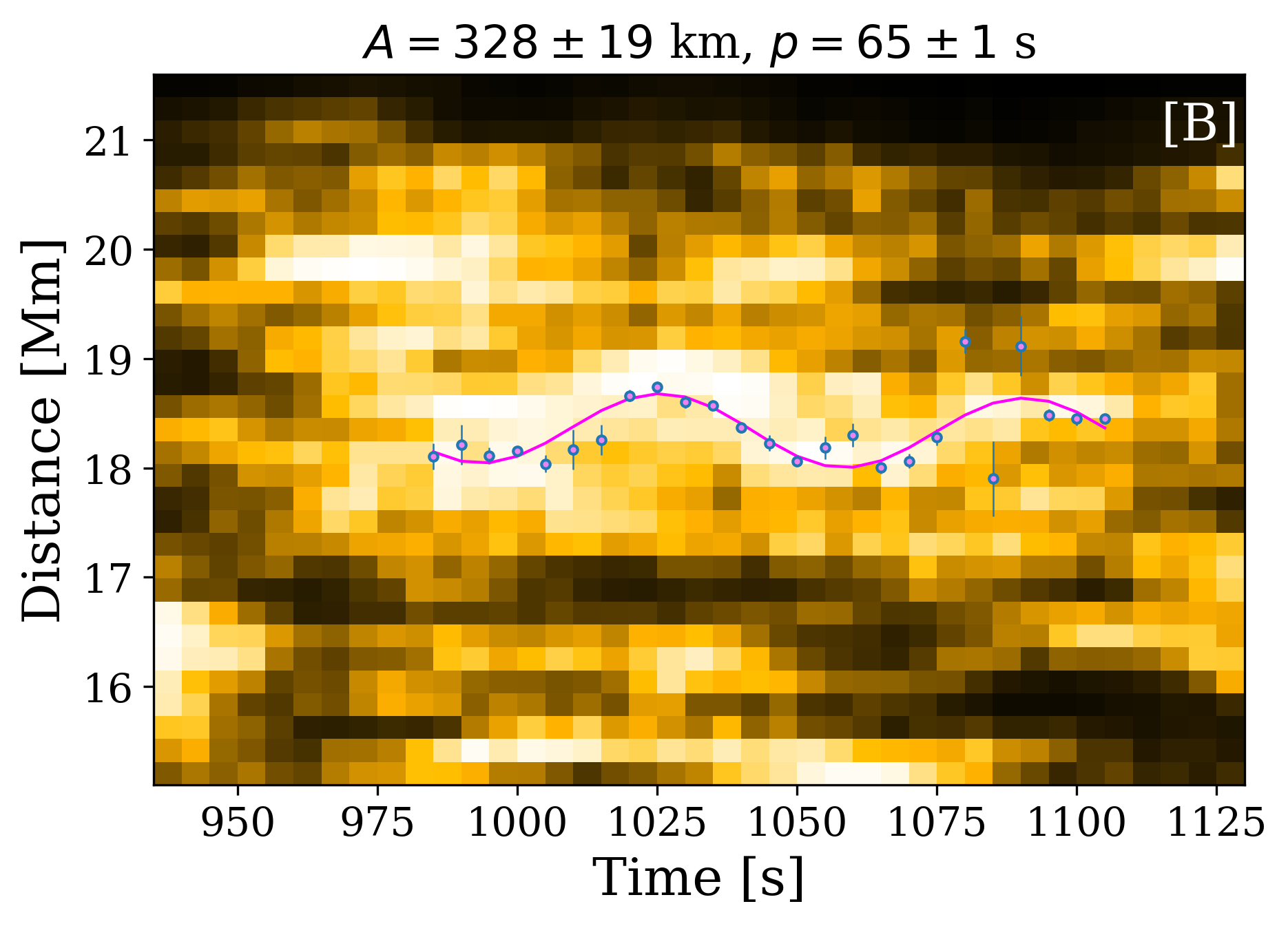}
    \includegraphics[width=0.40\textwidth]{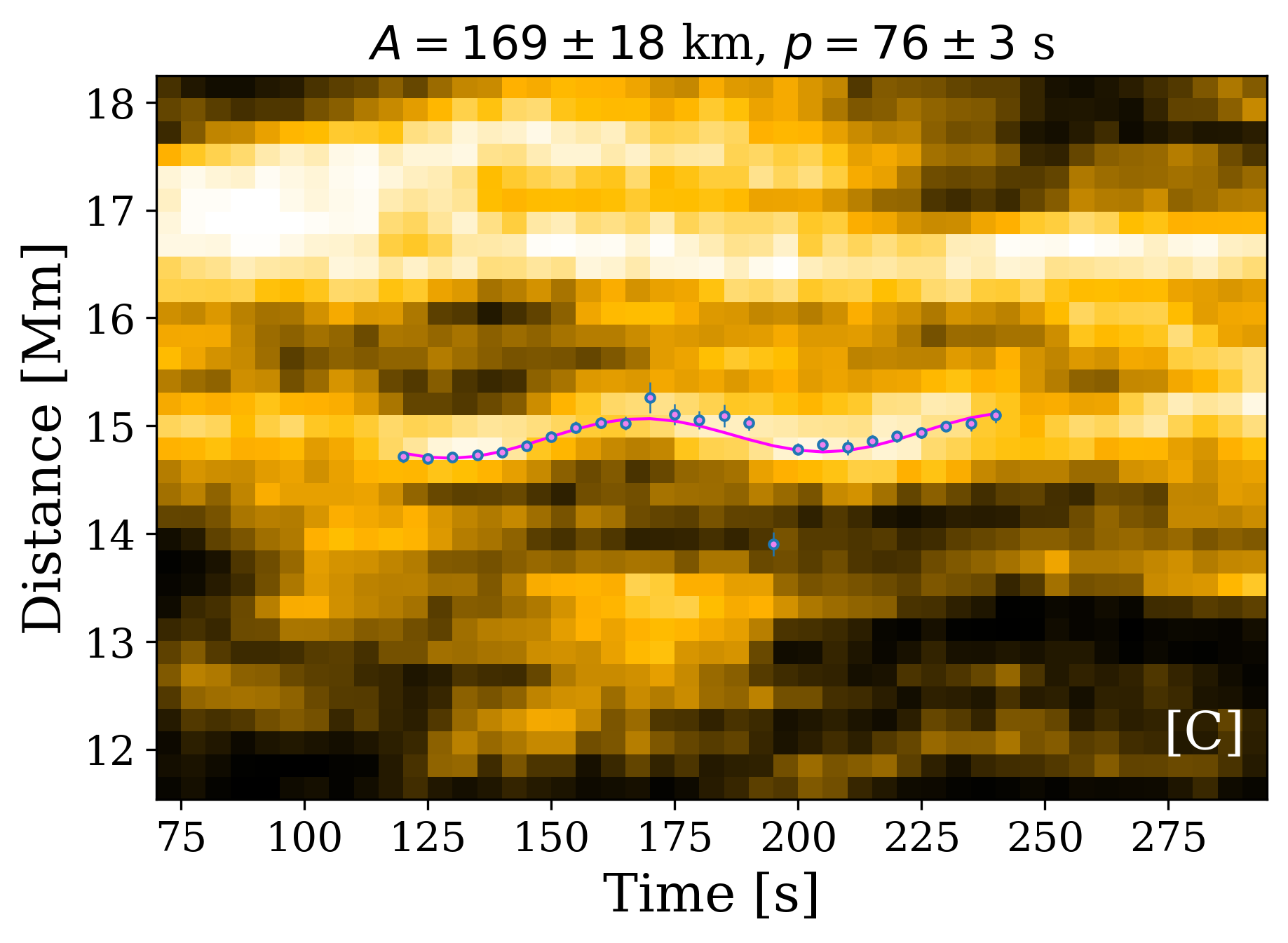}
    \includegraphics[width=0.40\textwidth]{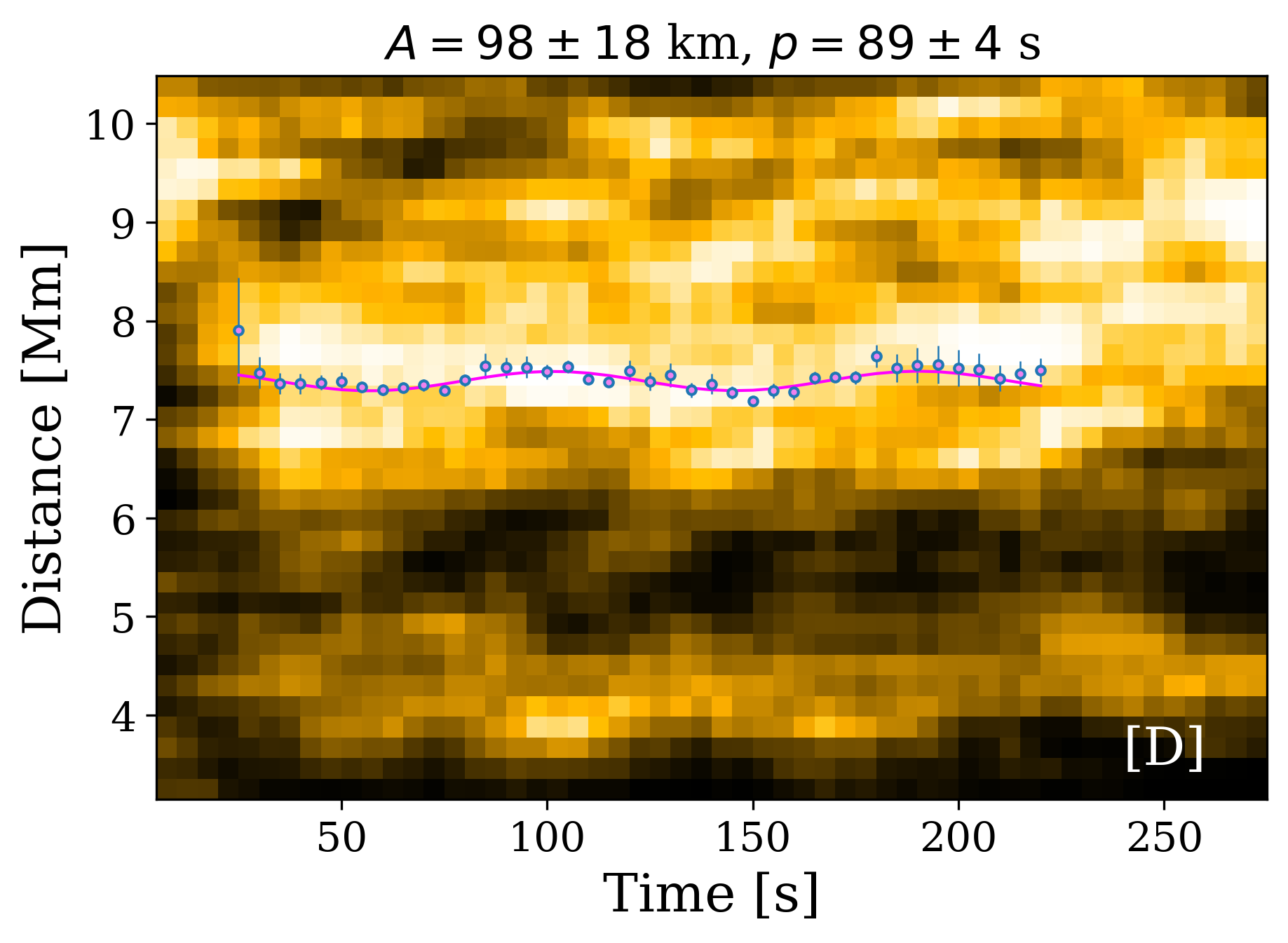}
    \includegraphics[width=0.40\textwidth]{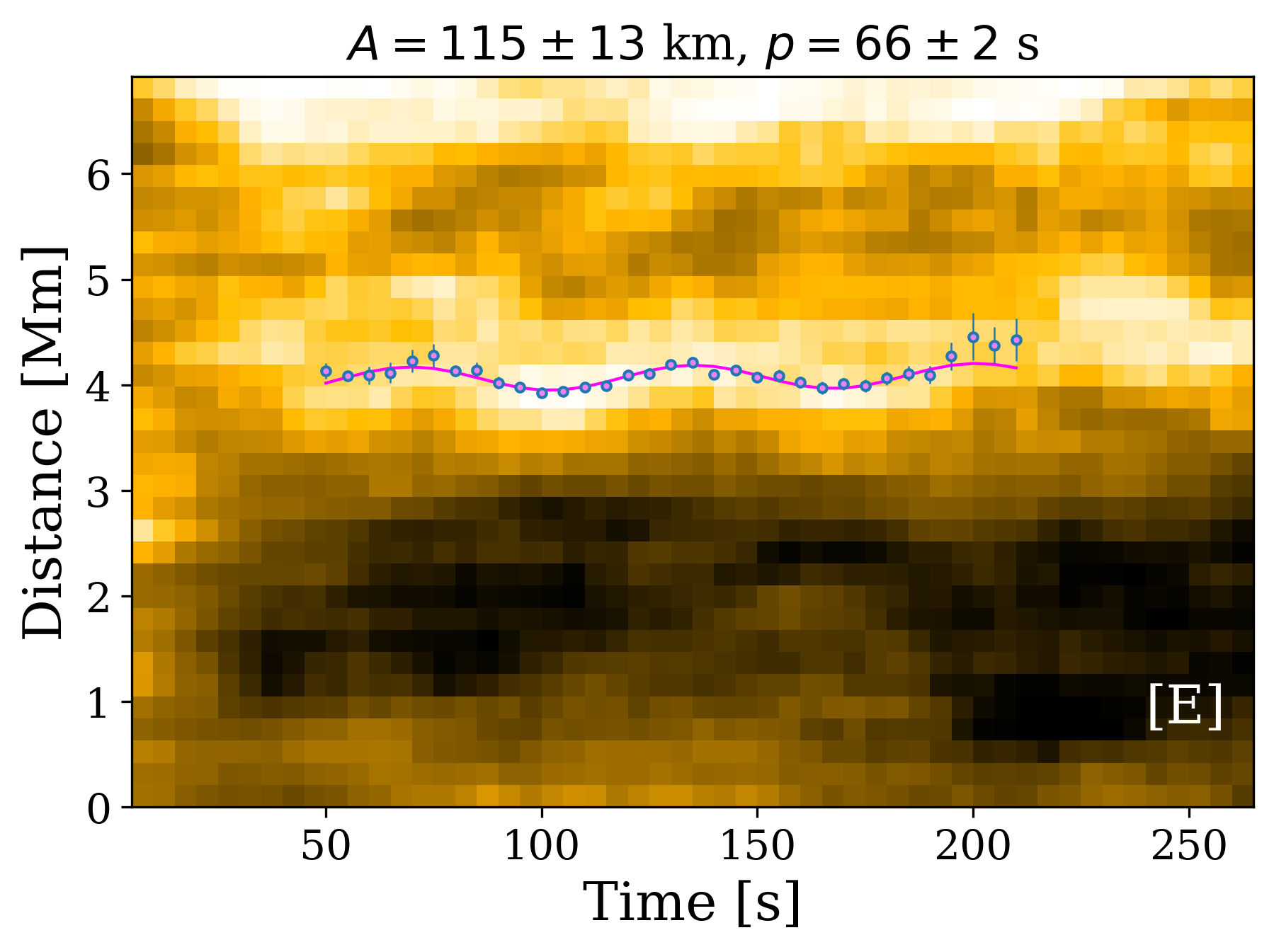}
    \caption{[A] The time-distance map for the transverse slit at h = 8.4 Mm in Slit 4. Dotted magenta curves correspond to the detected transverse waves fitted as mentioned in Section \ref{sec:fastmodedetection}. [B]-[E] Zoomed view of the visually detected transverse waves. The error bars represent the uncertainty on the mean of the fitted Gaussian at each time. The magenta curves show the corresponding fitted oscillations. The amplitude (A) and period (p) are mentioned for each fitted curve.} 
    \label{fig:transverse_oscillations_parameters}
\end{figure*}

To each transverse oscillation detected, a Gaussian function is fitted perpendicular to the oscillating structure to extract oscillation parameters, determining its center at each time step \citep{2024A&A...689A.295S}. The resulting center positions are then fitted with a sinusoidal function with a linear trend:
\[y = A \sin(2 \pi t/p + \phi) + Bt +C\]
where $A$ represents the displacement amplitude,
$p$ is the oscillation period,
$\phi$ is the phase,
$B$ and $C$ are constants representing the linear and constant background trend. This method is similar to the one used in \cite{2022A&A...666L...2M,2024Shrivastav, 2025Shrivastav} to fit the oscillations manually. 
This method is applied to transverse oscillations in TD maps of five plumes, though the number of transverse slits used varies between plumes. Figure \ref{fig:transverse_oscillations_parameters} represents one such example where panel [A] represents the transverse waves detected in slit 4 at a height of 8.4 Mm above the solar limb. The number of transverse waves at this height is only 4. The lower number of transverse waves at each height is because (i) we are considering the transverse waves in individual plume structures and (ii) these waves are fitted manually. Further, Panels [B]-[E] of the same figure represent the zoomed regions of the four fitted waves depicted by magenta dashed lines in Panel [A]. It should be mentioned that the Gaussian function is fitted at each time step, and the red circles represent the extracted local intensity maximum at each time step. The error bars correspond to the uncertainty in the mean of the fitted Gaussian at that time step and not one standard deviation. The magenta curve in each panel represents the fitted sinusoidal function to the local intensity maxima. The amplitude and the periodicity of the fitted oscillations are mentioned for each wave. The velocity amplitudes are calculated using \(v = 2 \pi A/p\), where $A$ is the displacement amplitude, and $p$ is the period. Further, only oscillations spanning at least one full cycle are selected, and transverse waves that can be distinctly isolated from the background are considered for analysis. A total of 98 transverse oscillations have been fitted in this way.

\begin{figure*}
    \centering
    \includegraphics[width=0.45\linewidth, height = 7.3cm]{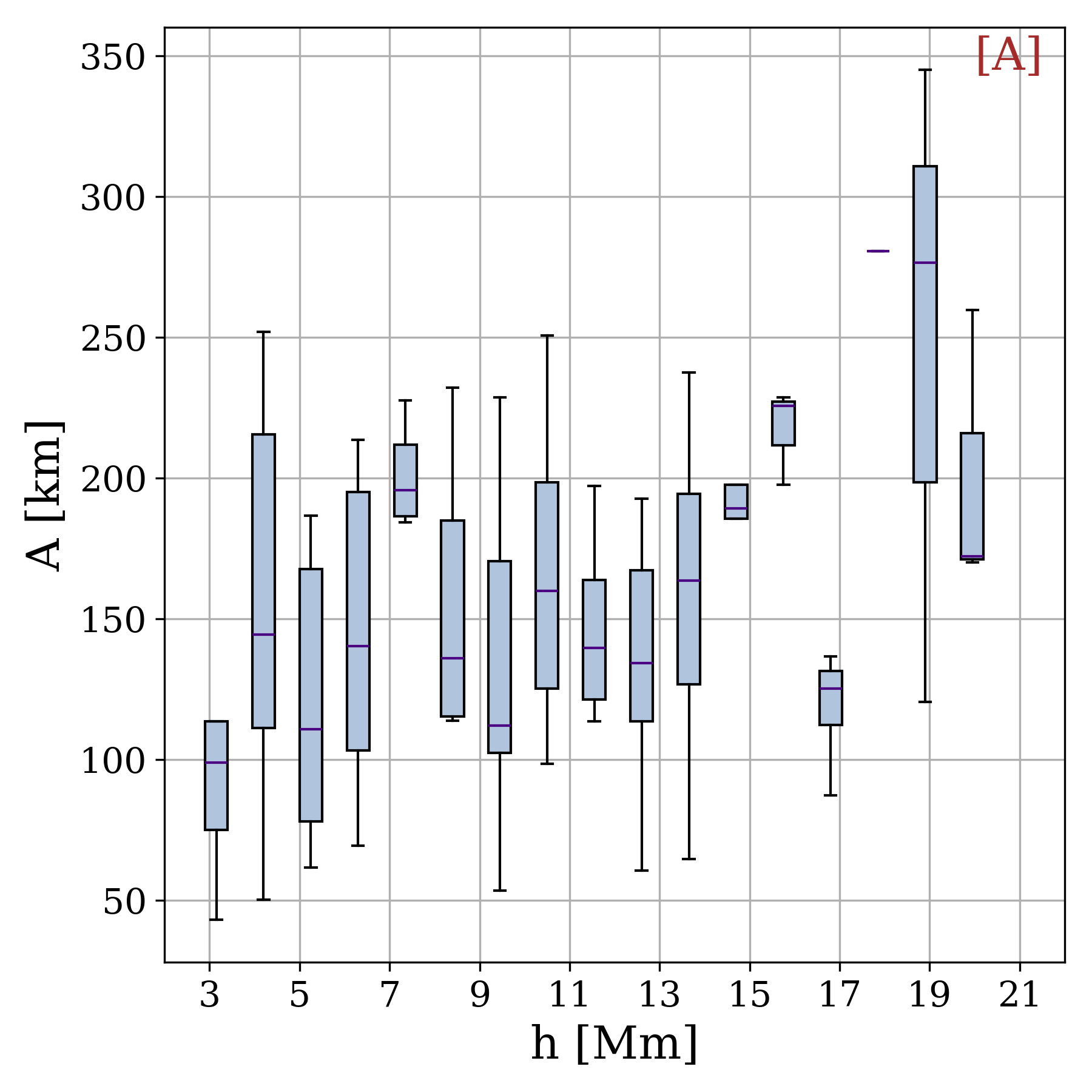}
    \includegraphics[width=0.45\linewidth, height = 7.3cm]{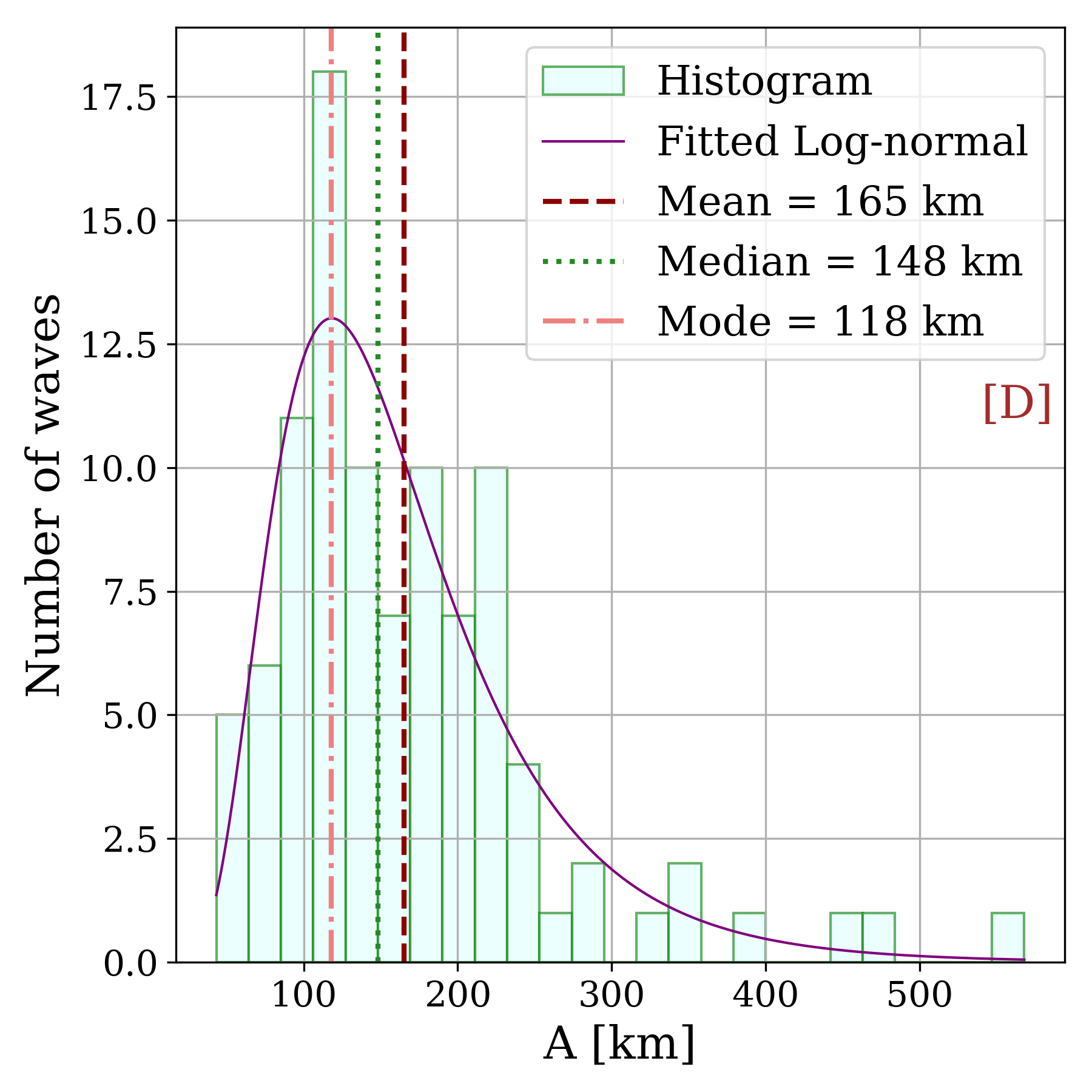}
    \includegraphics[width=0.45\linewidth, height = 7.3cm]{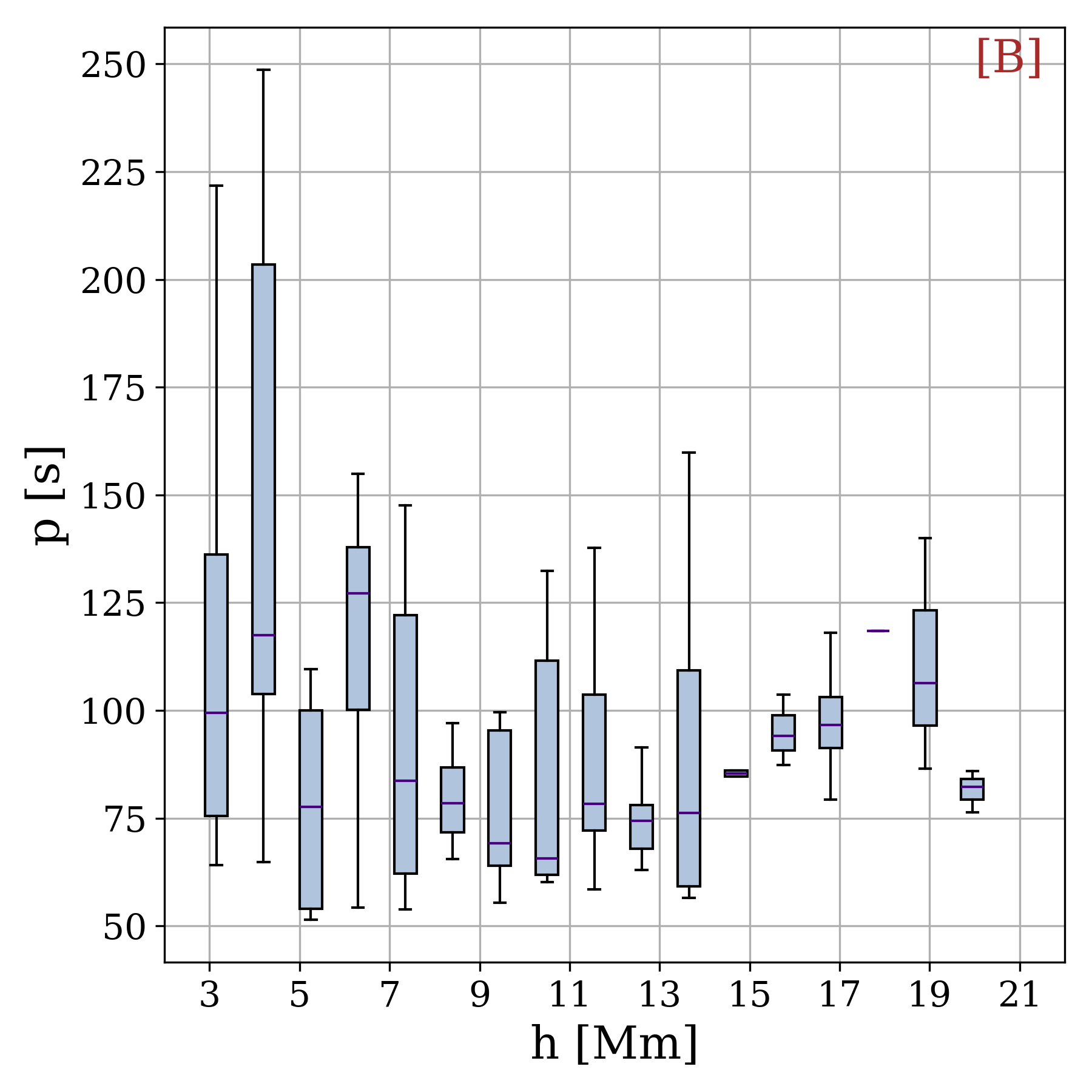}
    \includegraphics[width=0.45\linewidth, height = 7.3cm]{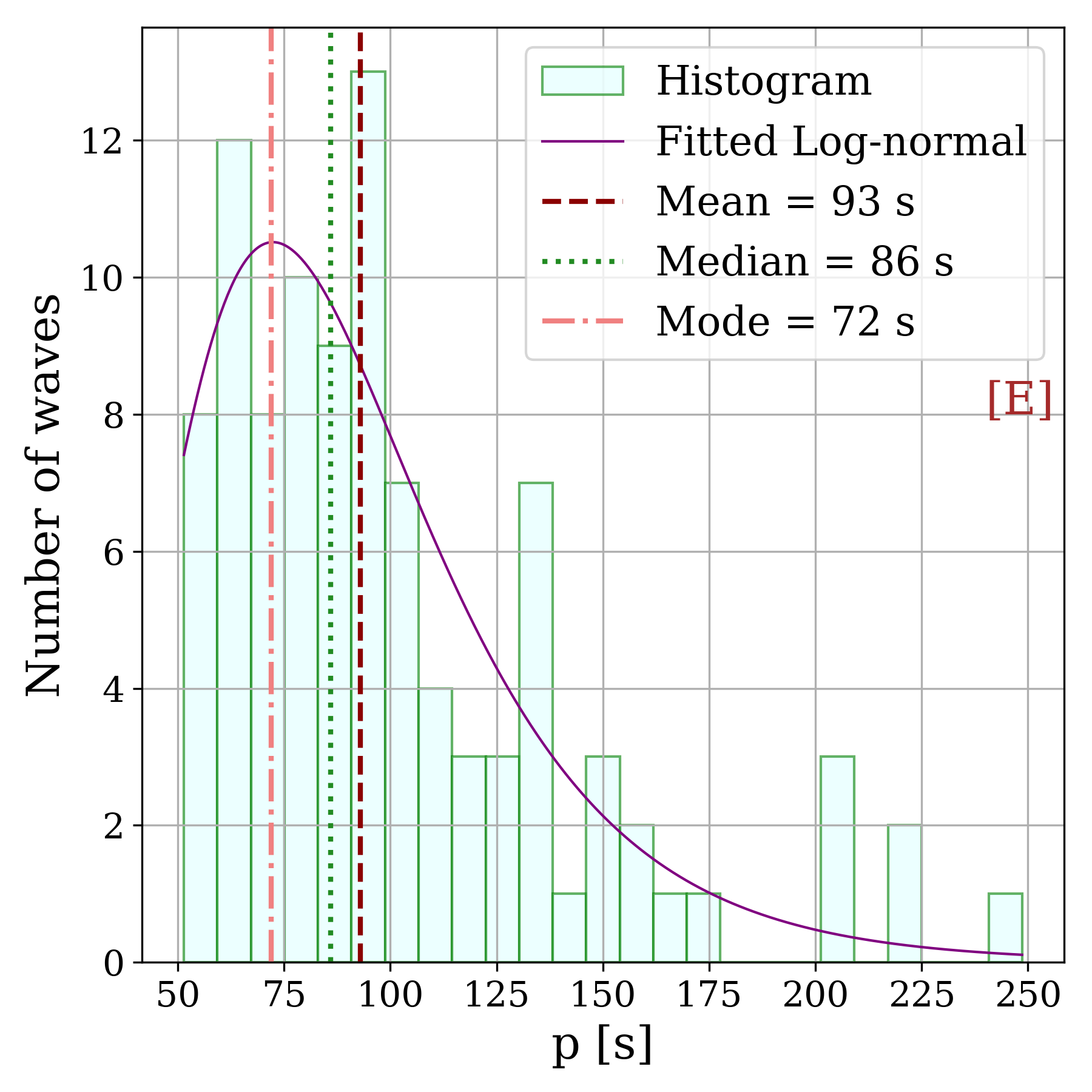}
    \includegraphics[width=0.45\linewidth, height = 7.3cm]{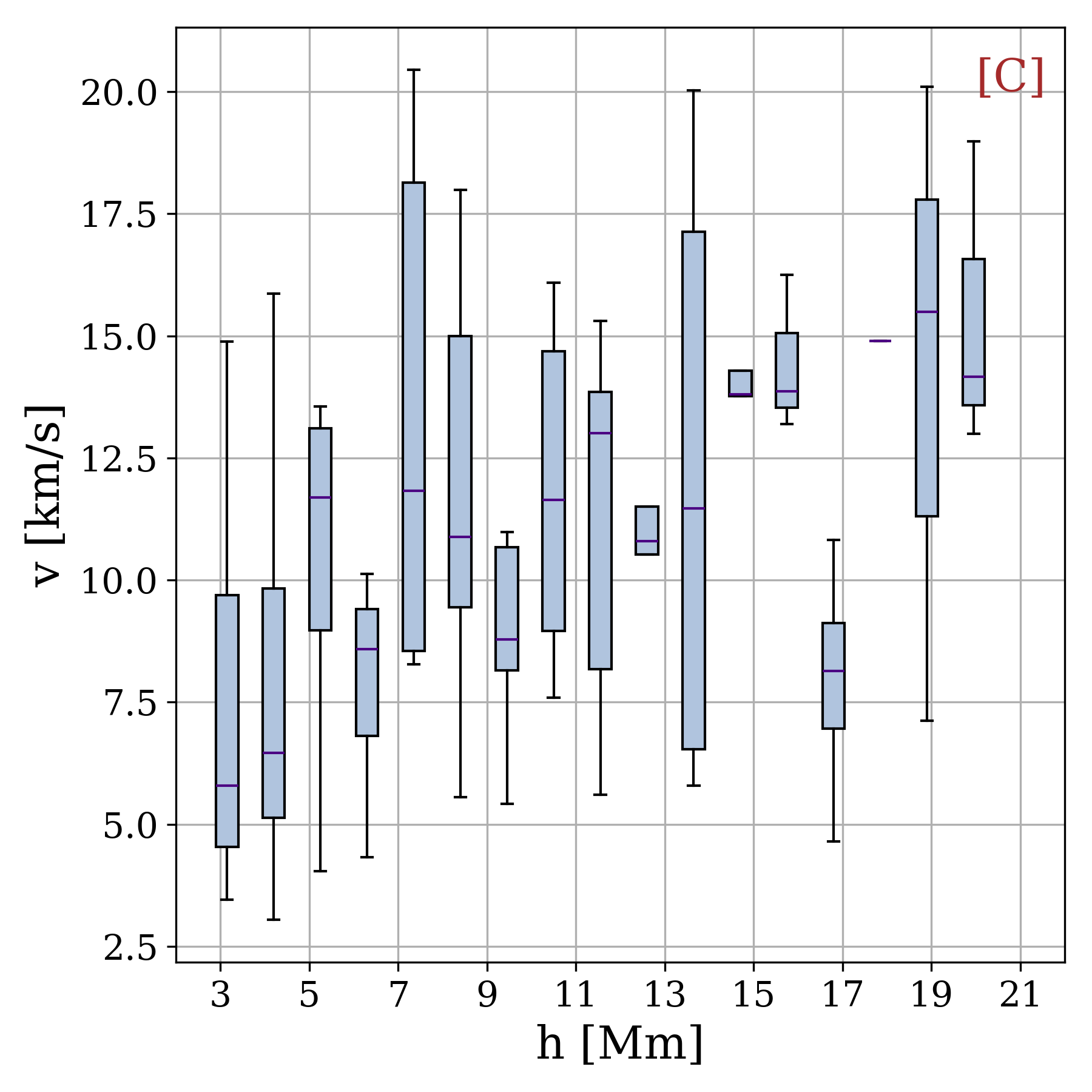}
    \includegraphics[width=0.45\linewidth, height = 7.3cm]{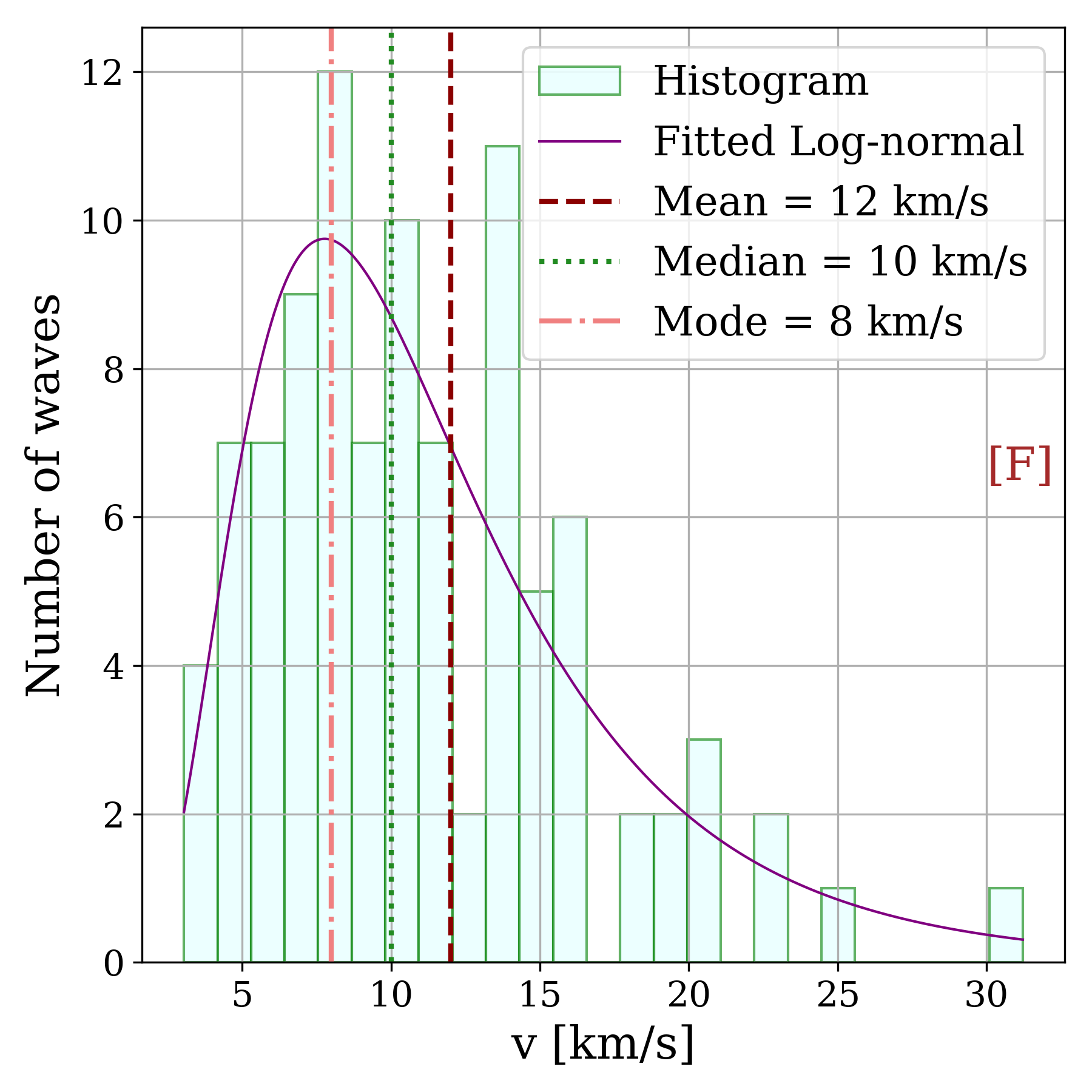}
    \caption{Panels [A] to [C] represent the box and whisker plot of the wave parameters obtained from all the selected plumes at different heights. Each box's upper and lower bounds correspond to the third and first data quartiles. The horizontal lines indicate the median values. Panels [D] to [F] histograms of the displacement amplitudes, periods, and velocity amplitudes for all the heights from 3 to 20 Mm above the solar limb for all the selected plumes. The purple curve represents the log-normal distributions obtained from the data using the arithmetic moments. The light coral, forest green, and dark red colored straight lines represent the mode, median, and mean values of the parameters obtained from the log-normal distributions.}
    \label{fig:box_whisker_plots}
\end{figure*}

\subsubsection{Wave parameter variation with height} \label{subssection:wave_parameters_transverse}
The variation of the wave parameters, namely displacement amplitudes $(A)$, periods $(p)$, and the velocity amplitudes $(v)$ of transverse waves, as a function of height above the limb $(h)$ is illustrated in Panel [A]-[C] of Figure \ref{fig:box_whisker_plots}. Despite the limited number of detected oscillations, there appears to be a trend of increasing displacement amplitude with height. This trend can be made robust by investigating a larger sample of plumes or utilizing a longer-duration dataset than the one under consideration. This similar behaviour is also reported earlier in polar coronal holes \citep{1998A&A...339..208B, 2015NatCo...6.7813M} and in coronal off-limb regions \citep{1998SoPh..181...91D} using non-thermal line widths. Looking at Panel [B] of the same Figure, it is clear that at long periods, waves are more prominent at lower heights, especially below 5 Mm.

The velocity amplitudes derived in this study are predominantly in the range of 2.5 to 20 kms$^{-1}$, whereas those reported using AIA/SDO extend up to 30 kms$^{-1}$ \citep{2020ApJ...894...79W,2025arXiv250113673M}. In addition, theoretical modelling suggests that these velocity amplitudes increase with height before either levelling off \citep{2020ApJ...899....1P} or decreasing after 45 Mm \citep{2024A&A...689A.195G}. The displacement amplitudes $A$ measured in this study range from 50 to 350 km within a height interval of 3 to 20 Mm, while in \cite{2020ApJ...894...79W}, $A$ spans from 200 to 1100 km for heights between 5 to 20 Mm. Similarly, the periods observed here range from 50 to 250 s, compared to 100 to 650 s reported in \cite{2020ApJ...894...79W}. These discrepancies can be attributed to the fine-scale features observed due to the high spatial and temporal resolution dataset obtained from the Solar Orbiter. It is likely that previous observations using the AIA dataset with a resolution coarser than HRI were unable to detect the lower period oscillations that have smaller amplitudes due to the limitations of temporal and spatial resolutions \citep{2023ApJ...952L..15L}. 

As the distance from the limb ($h$) increases, the number of detected transverse waves at each height decreases, likely due to increased noise with height or increased integration of intensity along LOS \citep{2019ApJ...881...95P}. Consequently, we are unable to detect transverse waves beyond 20 Mm. Moreover, the integration of intensity along LOS also forbids us to track individual transverse waves at different heights, and thus restricts us from direct estimation of their phase speed from imaging observations alone.
% Perhaps, this is the reason that the phase speed of transverse waves has not been estimated in \cite{2014ApJ...790L...2T, 2020ApJ...894...79W}.}

Further, despite the relatively lesser number of detected oscillations at each height compared to studies like \cite{2014ApJ...790L...2T,2018ApJ...852...57W,2020ApJ...894...79W} where automated detection methods were used, there is a significant spread in the distribution of wave parameters, underscoring the variability in wave behaviour across different heights. Due to the limited number of detected oscillations and considerable spread in the detected parameters at each height, we assessed the robustness of observed trends by performing the Spearman rank correlation analysis on the median values of parameters $A$, $p$ and $v$. The analysis reveals a statistically significant monotonic increase in both $A$ ($\rho_A$ = 0.62, probability = 0.008) and $v$ ($\rho_v$ = 0.69, probability = 0.002) with height. In contrast, the Spearman rank correlation analysis for $p$ does not show meaningful dependence on height.
% , however, panel [B] of Figure \ref{fig:box_whisker_plots} reveals some height dependent variation of periodicity $p$ with longer period transverse waves detected at the lower heights, and the shorter period oscillations more frequent at higher heights.
This height-dependent variation of $p$ is consistent with the findings of \cite{2020ApJ...894...79W}.
\subsubsection{Bulk properties of transverse oscillations}
The statistical distributions of different properties of the transverse oscillations are presented in panels [D]- [F] of Figure \ref{fig:box_whisker_plots}. The histograms shown for $A$ (Panel [D]), $p$ (Panel [E]), and $v$ (Panel [F]) are compiled across all the heights and slits/plumes under study. The key statistical ranges deduced from these histograms are as follows:
\begin{itemize}
    \item Displacement amplitudes ($A$) range between 50 and 600 km,
    \item Periods ($p$) vary from 50 to 250 seconds,
     \item Velocity amplitudes ($v$) lie between 3 and 32 kms$^{-1}$. 
\end{itemize}
On the contrary, \cite{2020ApJ...894...79W} reported transverse log-normal mean values of displacement amplitudes with log-normal means ranging from 484 km at 5 Mm to 763 km at 28 Mm and wave periods decreasing from 373 s at 5 Mm to 279 s at 33 Mm. 

These histograms for all these parameters $A$, $p$, and $v$ exhibit positive skewness with extended tails. To compare our results with previous studies \citep{2011Natur.475..477M,2014ApJ...790L...2T, 2020ApJ...894...79W}, we fitted the histograms with log-normal distributions, using arithmetic moments. The fitted log-normal curves are shown in purple in Panels [D]- [F] of Figure \ref{fig:box_whisker_plots}, yield the log-normal means, and standard errors of 165 $\pm$ 82 km for $A$, 93 $\pm$ 39 s for $p$, and 12 $\pm$ 7 kms$^{-1}$ for $v$, respectively. 

\subsubsection{Relative density variation with height}
Plumes that are over-dense plasma regions embedded in the more tenuous background, the transverse waves observed within them can be considered as kink modes \citep{2014ApJ...790L...2T, 2020ApJ...894...79W}. The speed of kink mode $(c_k^2(z))$ is given by 
\[c_k^2(z) = \frac{<B^2(z)>}{\mu_0 <\rho>}\]
where $\mu_0$ is the permeability of the free space, $<B^2(z)>$ represents the average squared magnetic field, and $<\rho>$ is the average density taken over the flux tubes and the ambient plasma. Furthermore, \cite{2001A&A...374L...9M} suggested that if the total energy flux remains conserved as waves propagate outwards, the velocity amplitude (\(v\)) will have the following dependence on the magnetic field strength, density, and Area (A):
\(v \propto \rho^{-1/4} (BA)^{-1/4}\). This assumption of conserved energy flux remains valid only in the case of ideal, dissipationless plasma and does not hold in realistic dissipative environments.
In testing the wave hypothesis, \cite{1998A&A...339..208B, 2020ApJ...894...79W,2024A&A...689A.195G} assumed constant $BA$ with height, simplifying the relation between the velocity amplitude and number density as:
\(v \propto n_e^{-1/4}\), where $n_e$ is the electron density. Such wave-based diagnostic of density has also been shown to be reliable under coronal conditions in numerical simulations by \cite{2018ApJ...856..144M}.

\begin{figure}
    \centering
    \includegraphics[width=\linewidth]{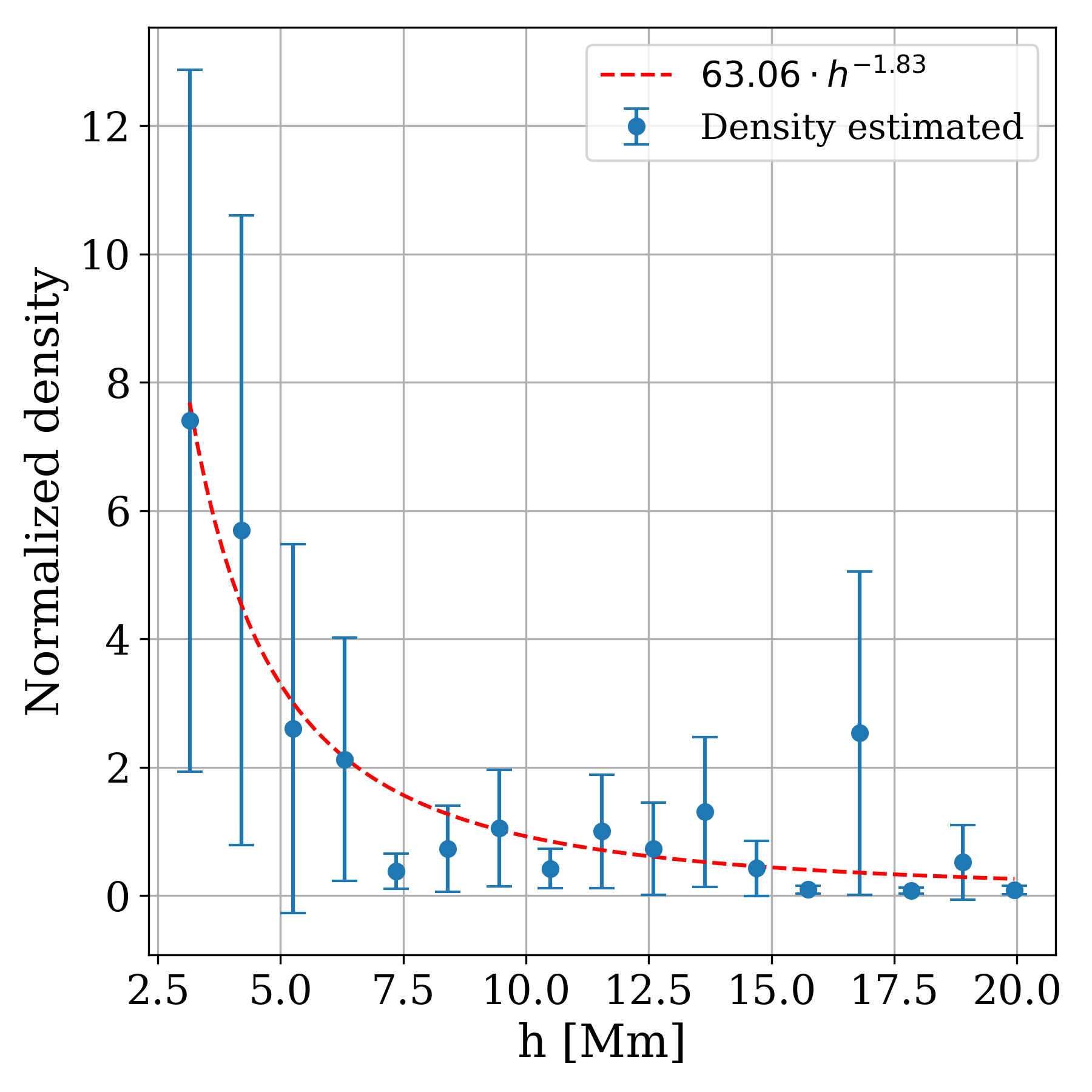}
    \caption{Relative variation in electron density with height obtained from observations in coronal plumes. Data has been normalized by their value at $\approx$12 Mm. The red dashed line corresponds to the polynomial fit. }
    \label{fig:relative_density_profile}
\end{figure}

Using the observed velocity amplitudes ($v$) of individual transverse oscillations, we derived the corresponding electron number density ($n_e$). Considering the mean values at each height, a relative density profile is constructed as shown in Figure \ref{fig:relative_density_profile}. Here, we have normalized the observed mean values w.r.t. the values measured at height 12 Mm. The error bars here represent the scatter of density values at that height scaled with the mean density at 12 Mm. Fitting this resulting profile with the power law equation using a least square fitting algorithm yields the following best-fit relation:

\[\frac{n_e(r)}{n_{e,o}} = 63.06h^{-1.83},\]
where $n_{e,o}$ corresponds to the electron density at height 12 Mm (following \citep{2020ApJ...894...79W}). The measured density profile is from 3 to 20 Mm above the solar limb, indicating a rapid decrease in the density with height as compared to that obtained in the previous studies by \cite{2012ApJ...751..110B,2020ApJ...894...79W}.
This rapid decrease in density with height represents the region where the solar atmosphere changes from chromosphere to corona. Assuming no dissipation is happening in this region, the scale height of the density profile between 3 Mm and 20 Mm is approximately 3 Mm. 

\section{Summary and Discussion} \label{sec:Summary}
In this letter, using the high-resolution EUI observations onboard Solar Orbiter, we present the first simultaneous observations of the co-existence of slow magneto-acoustic waves and transverse waves in the polar plumes. Previous works have either detected the simultaneous co-existence in plumes using different instruments \citep{2015ApJ...806..273L} or in fan loop structures \citep{2013A&A...556A.124T}. However, a detailed description of the wave parameters of both slow magneto-acoustic waves and Alfvénic waves obtained simultaneously within the same magnetic structures using the same instrument has not been performed. This study presents the first ever use of Solar Orbiter HRIEUI data for the detection of both slow magneto-acoustic and Alfvénic waves within the same magnetic structures. The main findings of our study are summarized as follows:
\begin{enumerate}
\item This study confirms the presence of both slow magneto-acoustic waves and Alfvénic waves co-spatially as well as co-temporally in the polar plumes, using the high-resolution observations of HRIEUV onboard Solar Orbiter.
\item These high-resolution observations enable us to find propagating disturbances in the plumes as well as in their fine-scale substructures. The propagating speeds corresponding to both regions are below the acoustic speed corresponding to the ambient temperature. Interestingly, even when wider slits are not able to reveal these disturbances within the plumes, they remain detectable within the fine-scale substructures. It is interesting to note that the identification of propagating disturbances within the fine-scale substructures in the plumes has not been reported earlier. The high-resolution dataset of HRIEUV made it possible to detect these fine-scale substructures. We suspect that the dataset that has higher resolution and better cadence than this dataset can further shed more light on the propagating disturbances in the fine-scale substructures in plumes and off-limb open loops. It may be noted that although we tried to isolate these fine-scale substructures in these plumes, however, sometimes, it became difficult to properly isolate these fine-scale substructures, and a broader slit was selected. Additionally, the appearance of the ridges at the same time in both the full plumes and the fine-scale substructures indicates the involvement of a coherent physical mechanism.
\item The amplitudes of the detected slow magneto-acoustic waves range from 1.4 to 3.2$\%$ of the background intensity, which is similar to the one reported in the previous studies \citep{2006A&A...448..763M,2013A&A...556A.124T}. However, the damping length of these waves estimated in this study ranges between 2 and 7 Mm, which is less than previously reported in the plumes using AIA datasets \citep{2014ApJ...789..118K, 2018ApJ...853..134M}. The reason for the smaller damping lengths is not yet clear. 
\item Due to the short duration of the dataset under consideration (19 min), periodicities of 9.4 min are detected in two selected plumes only along with shorter periodicities of 1.8, 2.6, and 6.25 min, as also reported in earlier studies \citep{2014ApJ...789..118K, 2018ApJ...853..134M}. 
\item The electron temperature in coronal plumes is estimated to range between 0.58 to 0.69 MK, which is consistent with the values reported earlier \citep{1998SSRv...85..371W,1998A&A...336L..90D,1999JGR...104.9753D,2003A&A...398..743D,2006A&A...455..697W,2008ApJ...685.1270L}
\item For the detection of the transverse waves, this study focuses exclusively on the region within coronal plumes, unlike previous studies that analyzed transverse waves using wide transverse slits across entire coronal holes \citep{2018ApJ...852...57W, 2020ApJ...894...79W,2025arXiv250113673M} or longer slits covering much of the polar region \citep{2014ApJ...790L...2T}. 
\item No automated detection technique is utilized here, instead 98 transverse oscillations are fitted manually within five plumes covering heights from $\approx$ 3 to 20 Mm. Due to the manual detection of these transverse waves, some small amplitude transverse oscillations may remain undetected in the procedure, thus keeping the method biased towards distinctly visible oscillations only.
\item The distributions of these wave parameters have lognormal means and standard errors for $A$, $p$, and $v$ are 174$\pm$91 km, 88$\pm$31 s and 12$\pm$7 kms$^{-1}$, respectively. Both the displacement amplitude and periodicity are smaller than the ones reported in the previous studies \citep{2014ApJ...790L...2T,2015ApJ...806..273L,2018ApJ...852...57W,2020ApJ...894...79W}. However, this distribution can be further improved by investigating more plumes or considering longer-duration datasets. 
\item The density scale height derived in this study is 3 Mm, which indicates that there is a sharp change in the density at 6 Mm above the solar limb, which is in contrast to the previously reported scale height in \citep{1998A&A...339..208B,2020ApJ...894...79W}.
% \item The estimated energy flux incorporating the density filling factor is 0.7 to 4.8 W m$^{-2}$, consistent with the previously reported values by \cite{2014ApJ...790L...2T} (recalculated in \cite{2014ApJ...795...18V}), confirming that the transverse waves do not have enough energy to accelerate the fast solar wind in open coronal structures, contradicting the conclusions by \cite{2011Natur.475..477M}.
\end{enumerate}

\begin{acknowledgments}
\textit{Solar Orbiter is a space mission of international collaboration between ESA and NASA, operated by ESA. The EUI instrument was built by CSL, IAS, MPS, MSSL/UCL, PMOD/WRC, ROB, LCF/IO with funding from the Belgian Federal Science Policy Office (BELSPO/PRODEX PEA 4000134088, 4000112292 and 4000106864); the Centre National d’Etudes Spatiales (CNES); the UK Space Agency (UKSA); the Bundesministerium für Wirtschaft und Energie (BMWi) through the Deutsches Zentrum für Luft- und Raumfahrt (DLR); and the Swiss Space Office (SSO). UB acknowledges the support from EUI guest investigatorship. UB would like to thank Cis Verbeeck, Nancy Narang and David Berghmans for their hospitality during her stay at ROB. V.P. is supported by SERB start-up research grant (File no. SRG/2022/001687). SKP is grateful to SERB/ANRF for a startup research grant (No. SRG/2023/002623). NN is supported by Belgian Federal Science Policy Office (BELSPO) through the contract B2/223/P1/CLOSE-UP. CV thanks BELSPO for the provision of financial support in the framework of the PRODEX Programme of the European Space Agency (ESA) under contract number 4000134088. TVD was supported by the C1 grant TRACEspace of Internal Funds KU Leuven, and a Senior Research Project (G088021N) of the FWO Vlaanderen. Furthermore, TVD received financial support from the Flemish Government under the long-term structural Methusalem funding program, project SOUL: Stellar evolution in full glory, grant METH/24/012 at KU Leuven. It is also part of the DynaSun project and has thus received funding under the Horizon Europe programme of the European Union under grant agreement (no. 101131534). Views and opinions expressed are however those of the author(s) only and do not necessarily reflect those of the European Union and therefore the European Union cannot be held responsible for them. TVD would like to thank Vaibhav Pant and Krishna Prasad for their hospitality during his sabbatical stay at ARIES in spring 2025.
The authors also acknowledge Dipankar Banerjee, Daye Lim, and the EUI team for their constructive comments.}
Finally, the authors wish to thank the reviewers for their valuable suggestions.
\end{acknowledgments}
\textit{Facilities:} Solar Orbiter 
\textit{Softwares:} SolarSoftware (SSW) \citep{1998SoPh..182..497F}, Python \citep{10.5555/1593511}

\appendix
\section{Slow magneto-acoustic waves in fine-scale substructures}\label{AppendixA}
\begin{figure*}
    \centering
    \includegraphics[width=0.20\linewidth]{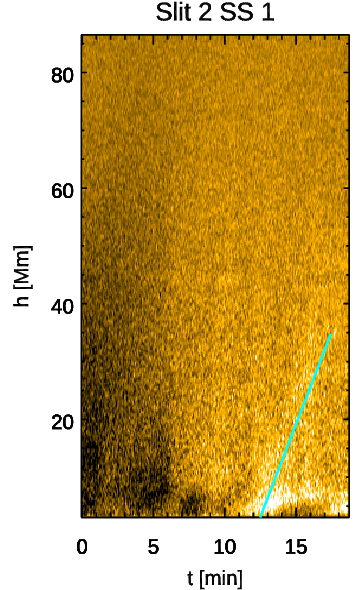}
    \includegraphics[width=0.20\linewidth]{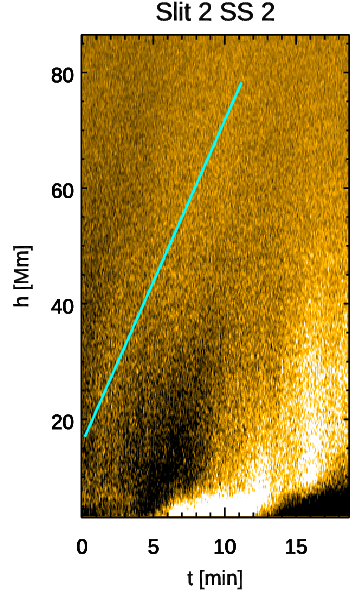}
    \includegraphics[width=0.20\linewidth]{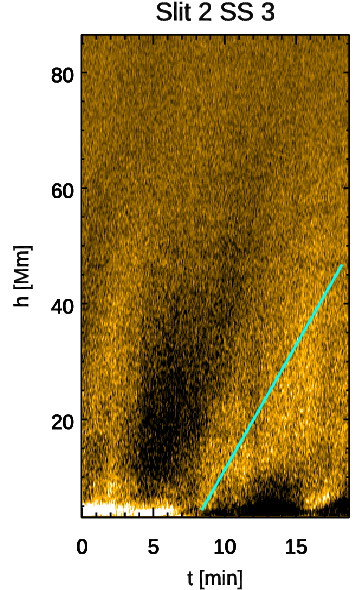}
    \includegraphics[width=0.20\linewidth]{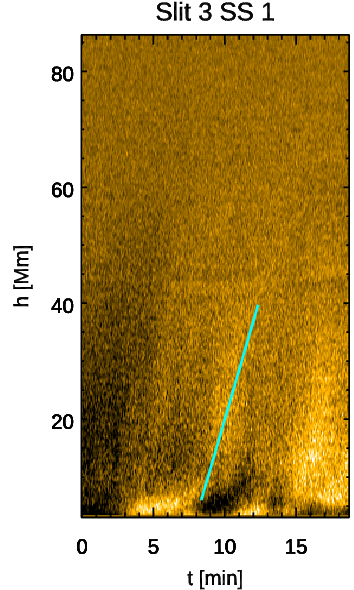}
    \includegraphics[width=0.20\linewidth]{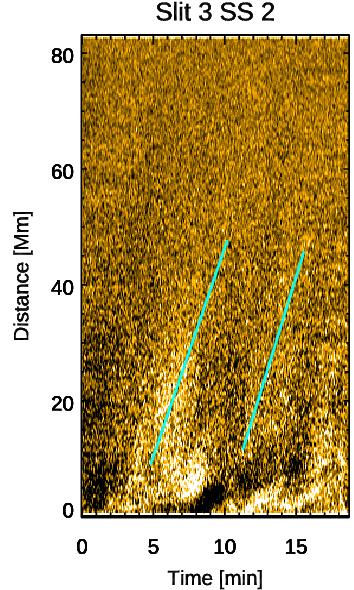}
    \includegraphics[width=0.20\linewidth]{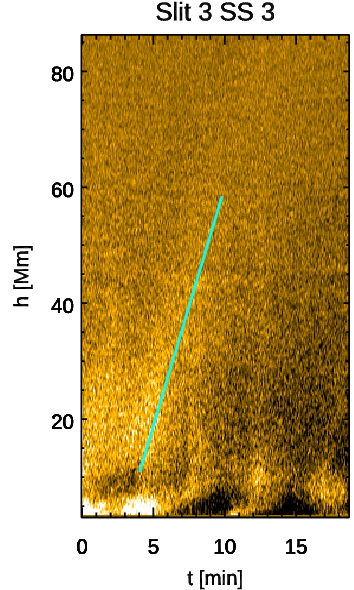}
    \includegraphics[width=0.20\linewidth]{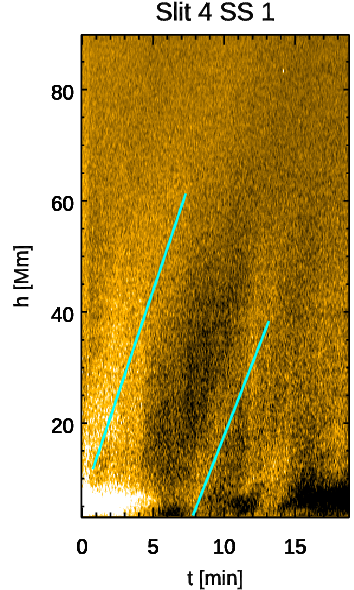}
    \caption{TD maps of the fine-scale substructures in Slits 2,3,4. Slit numbers are mentioned at the top of each TD map. The cyan line depicts the speed of the propagating disturbances.}
    \label{fig:vertical_slits_2to4}
\end{figure*}

For slit 2 and slit 3, there are three fine-scale substructures as shown in Figure \ref{fig:datasets} panel [C]. In the TD map of full slit 2, the ridge is not distinctly visible, still a cyan line indicating the ridge is marked in Figure \ref{fig:vertical_slits}. However, while examining the TD maps of its substructures, the ridges become identifiable in all three TD maps (Slit 1 substructure 1, substructure 2, substructure 3) as illustrated in Figure \ref{fig:vertical_slits_2to4}. For substructure 1 of slit 2, the intensity along the whole slit is uniform, resulting in a vertical ridge during 5-10 minutes of the TD map. In substructure 2, there appears a ridge initially, which is fitted with the cyan line. The subsequent ridges appearing at later times are because of the brightening happening at the base of the substructures. Only one single ridge is highlighted for substructure 3. It should be noted that a wider slit is used to fit this substructure, and the initial vertical bright ridge likely traces the plasma ejection happening. This substructure is composed of at least three unresolved fine-scale substructures, prompting the use of a wider slit for SS 3. It should be noted that the vertical bright ridge appearing in the Slit 2 substructure 3 (at time 0-3 minutes in TD map), and the fitted ridge of Slit 2 substructure 2 are clearly visible in the TD map of the whole slit. The ridges fitted in the TD map of substructures 1 and 3 correspond to the ridge fitted with a cyan line in the whole slit TD map. 
% From 10 to 17 minutes, there appears a bright ridge which is further composed of narrow vertical bright ridges, indicating the emergence of fine-scale dynamics within the substructure.
For slit 3, Substructure 1 (SS 1) and Substructure 3, there is one fitted ridge. In substructure 2 of the same plume, two ridges are distinctly identified and fitted with cyan lines, followed by the appearance of the bright vertical ridge at the end of the TD map. But for the two substructures 2 and 3, TD maps of the substructures corresponding to Slit 3, there appears a ridge which originates at a time less than 5 minutes, and the same ridge is also appearing in the TD map of the whole slit. This further confirms the presence of the slow magneto-acoustic waves in the plumes.

For Slit 4, while considering the whole slit, the ridge appeared to be discontinuous as mentioned earlier. However, while examining its fine-scale substructure, two distinct ridges are apparent. The first ridge fitted here in the SS was not distinctly visible in the TD map of the whole slit and was not fitted there. However, the second ridge fitted here corresponds exactly to the ridge appearing in the TD map of the whole slit. The enhanced brightness at the base of the TD map between times 12 to 15 minutes corresponds to the plasma ejection happening at the base of the slit (Solar X 80 to 100 arcsec). The average speed of these propagating disturbances observed in these fine-scale substructures is reported in Table \ref{tab:propagationspeed}.

\bibliography{bibliography}{}
\bibliographystyle{aasjournal}

\end{document}